\documentclass[lettersize, journal]{IEEEtran}

\usepackage{amsmath, amssymb, amsfonts, amsthm}
\usepackage{algorithm, algcompatible}
\usepackage{url}
\usepackage{array}
\usepackage{graphicx}

\usepackage{caption}
\usepackage{subcaption}

\usepackage{dblfloatfix}

\usepackage{adjustbox}
\usepackage{textcomp}
\usepackage{verbatim}
\usepackage{multirow}

\usepackage{cite}
\usepackage{color}
\usepackage{bm}     

\usepackage{acro}   

\makeatletter
\def\hlinewd#1{%
	\noalign{\ifnum0=`}\fi\hrule \@height #1 \futurelet
	\reserved@a\@xhline}
\makeatother

\newcolumntype{?}{!{\vrule width 1.5pt}}

\newtheorem{prop}{Proposition}

\newcommand{\specialcell}[2][c]{%
	\begin{tabular}[#1]{@{}c@{}}#2\end{tabular}}

\usepackage{pgfplots}
\usepackage{pgf}
\usepackage{tikz}

\usetikzlibrary{spy}

\usetikzlibrary{matrix}
\usetikzlibrary{decorations.pathmorphing} 
\usetikzlibrary{fit}					  
\usetikzlibrary{backgrounds}	          
\usetikzlibrary{shapes.arrows}
\usetikzlibrary{shapes.geometric}
\usetikzlibrary{shapes.multipart}
\usetikzlibrary{matrix, arrows, decorations.pathreplacing, positioning}
\tikzstyle{rec}=[draw,rectangle, minimum height=2cm]
\tikzset{>=stealth', punkt/.style={rectangle, 
		fill=gray!40, 
		draw=black, very thick, text width=3em, minimum height=2.5em, text centered}}
\tikzset{>=stealth', Denoi/.style={rectangle, fill=blue!20, 
		draw=black, very thick, text width=6em, minimum height=3.5em, text centered}}
\tikzset{>=stealth', CG/.style={rectangle, 
		fill=green!20, draw=black, very thick, text width=8em, minimum height=3.5em, text centered}}
\tikzstyle{background} = [rectangle, fill=green!20, inner sep=0.1cm, rounded corners=4mm, 
minimum height=6em]
\tikzstyle{sum}   = [draw, fill=gray!40, circle, node distance=1cm]
\tikzstyle{dot}   = [circle, fill=black, inner sep=0pt, minimum size=5pt, node contents={}]
\tikzstyle{fig_n} = [node distance=30pt, inner sep=0cm]

\def\*#1{\mathbf{#1}}
\def\+#1{\mathcal{#1}}
\def\-#1{\mathbb{#1}}
\def\~#1{\mathrm{#1}}
\def\@#1{\mathbb{#1}}
\def\R{\mathbb{R}}

\algnewcommand\INPUT{\item[\textbf{Input:}]}
\algnewcommand\OUTPUT{\item[\textbf{Output:}]}

\renewcommand{\hbar}{\bar{h}}

\newcommand{\transp}{^\top}

\begin{document}
	
\title{Rotational Augmented Noise2Inverse for Low-dose Computed Tomography Reconstruction}
	
\author{Hang Xu, Alessandro Perelli
        \thanks{This work involved human subjects in its research. The authors confirm that all human subject research procedures and protocols are exempt from review board approval. The AAPM low-dose CT dataset has obtained ethical approval by the local committee at Mayo Clinic USA. The dataset is made available through the Cancer Imaging Archive (TCIA) restricted license which require de-identifiability of the data and the authors have agreed with this requirement.}
		\thanks{H. Xu and A. Perelli are with the Department of Biomedical Engineering, University of Dundee, Scotland, DD1 4HN United Kingdom.}
		\thanks{\hspace*{-1.1em}Corresponding authors: A. Perelli, \mbox{\texttt{aperelli001@dundee.ac.uk}} }
}
	
\maketitle
	
\begin{abstract}
	In this work, we present a novel self-supervised method for Low Dose Computed Tomography (LDCT) reconstruction. Reducing the radiation dose to patients during a CT scan is a crucial challenge since the quality of the reconstruction highly degrades because of low photons or limited measurements. Supervised deep learning methods have shown the ability to remove noise in images but require accurate ground truth which can be obtained only by performing additional high-radiation CT scans. Therefore, we propose a novel self-supervised framework for LDCT, in which ground truth is not required for training the convolutional neural network (CNN). 
	Based on the Noise2Inverse (N2I) method, we enforce in the training loss the equivariant property of rotation transformation, which is induced by the CT imaging system, to improve the quality of the CT image in a lower dose. 
	Numerical and experimental results show that the reconstruction accuracy of N2I with sparse views is degrading while the proposed rotational augmented Noise2Inverse (RAN2I) method keeps better image quality over a different range of sampling angles. Finally, the quantitative results demonstrate that RAN2I achieves higher image quality compared to N2I, and experimental results of RAN2I on real projection data show comparable performance to supervised learning. 
\end{abstract}
	

\section{Introduction}

\IEEEPARstart{C}{omputed} Tomography (CT) is routinely used in medical diagnostics and the amount of X-ray radiation dose constitutes a critical safety concern. 
The reconstruction of high-quality images of the inner part of the body from a set of low-dose CT (LDCT) measurements is a fundamental objective in medical imaging. 

LDCT can be achieved either by reducing the number of projection angles acquired during a CT scan or by decreasing the X-ray source intensity. LDCT image reconstruction is a challenging problem since employing sparse projection measurements leads to solving highly under-determined systems while low X-ray photons count induces high spatially dependent noise. 

The inverse problem of reconstructing the unknown image from LDCT measurements is based on ensuring data consistency by leveraging the physics knowledge of the CT data acquisition encoded in the forward model and a regularizer which impose the prior information on the expected image's structure \cite{venkatakrishnan2013plug}. Many regularizers have been proposed to date, including those based on low-rank penalty \cite{hu2011lowrank}, generalized Total Variation \cite{niu2014}, transform-domain sparsity \cite{danielyan2011bm3d} and dictionary learning \cite{elad2006dict}\cite{xu2020limited}.

Deep learning (DL) has gained enormous popularity in medical image reconstruction and particularly Convolutional Neural Networks (CNN) are widely deployed with state-of-the-art performance in image denoising, whose aim is to enhance a noisy image to its high-quality counterpart \cite{wang2020deep,wang2023review}. 

A popular approach is constituted by using post-processing techniques where the reconstructed image is obtained using a fast algorithm which ensure consistency and then the CNN is used to denoise and enhance the estimated image \cite{ongie2020deep}. 

Another class of methods refers to learned iterative techniques which are generally more accurate but require more computational resources and time \cite{lucas2018}. Furthermore, these networks are trained using supervised learning which involves a dataset consisting of pairs of input noisy reconstructed images and high-quality target reconstructions \cite{zhang2017beyond}. In clinical applications, this is a challenging constraint since it is not possible to directly access the high-quality images but only retrospectively. 

Although high dose CT images can be used as target for supervised learning with accurate reconstruction results \cite{chen2017low,kang2017deep,huang2022deep}, in practise this is a potential drawback since the acquisition of sufficient number of ground truth images is impractical as it requires high radiation dose scans and moreover the physical settings of the CT system used for training should match the ones used for testing. 

Recently unsupervised learning has sparked tremendous interest since these methods rely exclusively on the information available in the corrupted measurement data itself. Many deep learning image denoising methods have been developed where ground truth images are not required, providing crucial benefits in applications such as LDCT \cite{zeng2021noise,jeon2022mm}.

Noise2Noise \cite{lehtinen2018noise2noise} presented a training procedure which allows to use only noisy pair of images, instances of the same hidden clean image corrupted by a different amount of zero-mean noise. While Noise2Noise still requires pairs of noisy data from the same image, this framework led to a variety of new approaches for self-supervised learning. 

Noise2Self \cite{batson2019noise2self} proposed to generate pair of data from a single image by creating a partition on the noisy image using a masking scheme; non-adjacent pixels are extracted from a noisy image as the target and the remaining pixels are the input. Noise2Void \cite{krull2019noise2void} introduced an improvement in the mask scheme such that the receptive field of each pixel excludes the pixel itself in order to avoid network degeneration. A "blind-spot" architecture with receptive field restricted to a different direction has been proposed in \cite{laine2019high} to improve over Noise2Void approach for image denoising. However, this kind of self-supervised methods strictly assume that the noise is element-wise independent in the image domain, which does not hold for CT reconstructed images due to the mixing forward operator \cite{kang2017deep,hendriksen2020noise2inverse}. 

Noise2Inverse (N2I) \cite{hendriksen2020noise2inverse} aims to overcome the limitation of  Noise2Self for inverse problems by exploiting particular partitions of the measurements (sinogram) to compute multiple statistically independent reconstructions. However in the sparse view CT case, N2I reconstruction becomes heavily affected by streaking artefacts since the number of measurements is lower than the dimension of the input image space, hence the discrete physical forward operator has a non trivial nullspace and the filtered backprojection (FBP) cannot capture information outside the range of the operator. 

Recently the concept of Equivariant Imaging (EI) was introduced in \cite{chen2021equivariant} and afterwards proposed for image restoration \cite{chen2022robust} where equivariant transformations are used to train the inverse mapping from compressed CT measurements. EI is an elegant formulation to learn the missing information outside the range but the performance of EI degrades rapidly in the presence of measurement noise unless the Stein’s Unbiased Risk Estimator is used \cite{chen2022robust} whose formulation is complex and depends on the type of noise. The equivariant self-supervised learning (E-SSL) framework \cite{dangovski2021equivariant} was recently developed to encourage sensitivity to certain transformations within contrastive learning.

\subsection{Main Contribution}

We propose a new equivariant-based model in the image domain within the N2I training strategy, motivated by the LDCT scenario where the measurements are sparse and highly corrupted by Poisson noise. The developed self-supervised training strategy is based on a novel loss function obtained by adding in N2I training a finite group of equivariant transformations to improve the prediction outside the range of the CT forward operator. 

The intuition of the proposed work is that it is possible to learn the missing information in LDCT by integrating the equivariant property in the image domain of the N2I model, i.e. by imposing the prior information that the CT system is invariant to group of transformations such as rotations. 

The contributions are: 1) a new self-supervised method called Rotational Augmented Noise2Inverse (RAN2I) for LDCT image reconstruction which does no require ground truth training dataset; 2) improve the image quality, by using rotational augmentation in the training loss, compared to N2I and achieve similar results with supervised methods.

\subsection{Notation and Paper Organization}

We adopt the following notations throughout the manuscript: discrete operators or matrices and column vectors are written respectively in capital and normal boldface type, i.e. $\*A$ and $\*a$, to distinguish from scalars and continuous variables written in normal weight; $[\*a]_j$ represents the $j$-th entry of $\*a$; `$(\cdot)\transp$' denotes the transposition while $\vec{\cdot}$ represent a vector in a 2D space; an image $\*x\in\R^d$ is represented by a column vector for algebraic operations. Finally, the expectation respect to random variables $a,b$ is indicated with the notation $\@E_{a,b}$.

The structure of the paper is organized as follows: in section \ref{ch2:self_super} we introduce the CT image reconstruction problem. Section \ref{subsec:N2I} reviews the implementation of N2I. Section \ref{ch3_EN2I} describes our proposed RAN2I framework and section \ref{sec:Setups} introduces the common settings for both numerical simulation and read data experiment. Section \ref{sec: numerical results} presents the results of the numerical simulations and section \ref{sec: experimental results} shows experimental results compared with the supervised method. Section \ref{ch5_discuss} provides a discussion and future work.

\section{CT reconstruction problem}\label{ch2:self_super}

X-ray CT produces images of attenuation coefficients and we consider a monoenergetic X-ray source with a detector array recording the intensity of the radiation exiting from the object along a number of paths without scatter. If the intensity of the source of radiation, $I_0$, passing through the object is known, then Beer's law provides the expected intensity after transmission, $I_i$ of the $i$-th ray as $I_i=I_0 e^{-\int_{L_i} x({\vec\nu}) dl}$ where $\int_{L_i} \cdot dl$ is the line integral along $L_i$ which is the path of the $i$-th ray through the object from the source to the detector and $x(\vec{\nu})$ is the spatial distribution of attenuation. The collected photons $I_i$ follow the Poisson distribution 
\begin{equation}
I_i\sim \mathrm{Poisson}\left\{I_0 e^{-\int_{L_i} x(\vec{\nu}) dl}\right\}
\end{equation} 
for each angle $i=1,\ldots, K$.

To discretize the measurement, we use a forward operator $\*A$ to convert the continuous measurement $\int_{L_i} x(\vec{\nu}) dl\approx \sum_j^N\*a_{ij}\*x_j$; therefore the measurement (sinogram) $\*y\in\mathbb{R}^{M\times 1}$ after the post-log can be described by
\begin{equation}
	\*y = \*A\*x+\*\epsilon
\end{equation}
where $m = KN_{dec}$ with $N_{dec}$ the fixed number of detectors, $\*A=[a_{ij}]\in\mathbb{R}^{m\times d}$ is the forward operator, $\*x\in\mathbb{R}^{d\times 1}$ is the vectorized input image and $\*\epsilon\in\mathbb{R}^{m\times 1}$ is the additive Gaussian noise with diagonal covariance matrix that approximates the Poisson noise \cite{buzug2009computed}. To estimate the CT image $\hat{\*x}$ from $\*y$, the filtered backprojection (FBP) $\*R=\*A^T\*F$ can be used to approximate $\*A^{\dagger}$ in the noiseless case, where $\*A^T$ is the backprojection operator and $\*F$ is the frequency weighting which for all projections at the same view angle is identical and can be implemented using a Ram-Lak filter \cite{hansen2021computed}. In the noisy case, applying $\*R$ to the projections leads to 
\begin{equation}\label{eq:FBP_noisy}
	\hat{\*x}=\*R\*y = \*R\*A\*x + \*R\*\epsilon
\end{equation}
In the sparse view LDCT case, the FBP generally shows streaking artefacts since accurate estimation of $\*x$ is impossible when $m \ll d$ without imposing additional constraint. Moreover \cite{schwab2019deep} adding to the FBP any image belonging to the nullspace of $\*A$, this ensures the measurement consistency  $\*A\*R\*y=\*y$. The case of low photon counts $I_0$, with $m\gg d$, guarantees uniqueness of the solution  $\hat{\*x}$ but it causes higher amount of spatially dependent noise due to the term $\*R\*\epsilon$.

\section{Self-supervised Deep Learning for CT Reconstruction}\label{subsec:N2I}

We consider the CT acquisition scenario where we can obtain noisy measurements 
\begin{equation}\label{eq:noise_mod}
	\*y \sim \*y^*_i + \bm\epsilon,\quad i=1, \ldots, N
\end{equation}
with the noisy free term $\*y^* = \*A\*x_i$, from a set of unknown images $\*x_1, \ldots, \*x_N\sim \*x$  sampled from some random variable $\*x$ with the noise $\bm\epsilon$ element-wise independent and mean-zero conditional on the data
\begin{equation}\label{eq:expectation}
	\@E[\*A\*x + \bm\epsilon |\,\*A\*x = \*y] = \*y
\end{equation}

By employing the FBP algorithm, the reconstructed images can be obtained as in (\ref{eq:FBP_noisy}) $\hat{\*x}_i=\*R\*y_i = \*R\*A\*x_i + \*R\*\epsilon_i$. 

The objective of the N2I algorithm is reconstruct the image that would have been obtained in the absence of noise
\begin{equation}\label{eq:FBPnoisy}
	\*x^*_i = \*R\*y^*_i.
\end{equation}

The proposed self-supervised method is based on the concept of Noise2Noise (N2N), in which the training image pairs are noisy images only. If mean squared error ($L_2$ loss) is used to minimize the parameters $\theta$ of the neural network $f_{\*\theta}$, the loss function of N2N can be written as:
\begin{equation}\label{eq:loss_N2N}
	L_{N2N}(\*\theta) = \frac{1}{N}\sum_{i=1}^{N}\|f_{\*\theta}(\*R\*A\*x_i + \*R\*\epsilon_{i,1}) - (\*R\*A\*x_i + \*R\*\epsilon_{i,2}) \|^2_2
\end{equation}
If the noise is zero mean
\begin{equation}
	\@E\left[\*R\*\epsilon_{i,2} | \*R\*A\*x = \*R\*y^*\right] =0
\end{equation}
and $\*R\*\epsilon_{i,1}$, $\*R\*\epsilon_{i,2}$ are independent, then the equation (\ref{eq:loss_N2N}) is equivalent to the supervised learning optimization problem
\begin{equation}\label{eq:loss_sup}
	L_{sup}(\*\theta) = \frac{1}{N}\sum_{i=1}^{N}\|f_{\*\theta}(\*R\*A\*x_i + \*R\*\epsilon) - \*R\*A\*x_i \|^2_2
\end{equation}

However, N2N requires two same CT scans to construct the training dataset, which is difficult to obtain. If the single image is split to obtain two noisy images in the image domain, $\*R\*\epsilon_{i,1}$, $\*R\*\epsilon_{i,2}$ are not statistically independent due to the reconstruction function $\*R$. 

Despite in noise $\*R\*\epsilon_{i,1}$, $\*R\*\epsilon_{i,2}$ in the image domain is statistically dependent, the intuition of N2I is to partition the data in the measurement domain, where the noise is element-wise independent as in Eq. \eqref{eq:noise_mod}, and training the network in the reconstruction domain. The partition $\-J$ divides the measurement into a target section $J\in\-J$ and an input section $J^C$ which is the complement of $J$. 
The process of partition assumes that we have a stack of sinograms $\{\*y_1, \*y_2, \ldots, \*y_N\}$ acquired from $K$ projection angles $\{\phi_1,\phi_2,\ldots,\phi_K\}$. Each sinogram $\*y_i$, with $i = 1,\ldots, N$, is split into $S$ subsets such that $\*y_{i,j},\,j=1,\ldots, S$ subset contains pixels from angles $\{\phi_j, \phi_{j+S}, \phi_{j+2S}, \ldots, \phi_{j+K-S}\}$. 
The FBP is applied to each partitioned sub-sinograms, yielding sub-reconstructions $\hat{\*x}_{i,j}=\*R_j \*y_{i,j}$. Then the subsets are divided and combined into the target $J$ and the input  $J^C$. The training target for the CNN is computed as the mean of section $J$, i.e., $\hat{\*x}_{i,J}=\frac{1}{|J|} \sum_{j\in J}\hat{\*x}_{i,j}$ while the training input is the mean of $J^C$ 
\begin{equation}
	\hat{\*x}_{i,J^C}=\frac{1}{|J^C|} \sum_{j\in J^C}\hat{\*x}_{i,j}
\end{equation}

The number of splits $S$ and the combination strategy are free parameters \cite{hendriksen2020noise2inverse}. In the training, the optimal weights $\hat{\*\theta}$ of the network $f_{\*\theta}$ are computed by minimizing the following loss function
\begin{equation}\label{eq:opt_n2i}
	\hat{\*\theta} = \arg\min_{\*\theta}\sum_{i=1}^N\sum_{J\in\-J}\|f_{\*\theta}(\hat{\*x}_{i,J^C}) - \hat{\*x}_{i,J} \|_2^2
\end{equation}
with $N$ the number of training images. Eq. (\ref{eq:opt_n2i}) is the approximation of the regression function that minimizes the expected prediction error \cite{hastie2009elements}. After training, the CNN is applied to denoise the FBP input $\hat{\*x}_{i,J^C}$ with the pre-trained network $f_{\*\theta}$ to obtain the estimated solution $\*x^*$. 

The output is obtained by applying the trained network to each partitioned-based reconstruction and computing the average
\begin{equation}
	\*x^*_i = \frac{1}{|J|}\sum_{j\in J}f_{\*\theta}\left(\hat{\*x}_{i,J^C}\right)
\end{equation}

In our implementation of self-supervised learning, we split the noisy sinograms $\*y_i$ obtained from $K$ angles indexed by $\{0, 1, 2, \ldots, K-1\}$ into two subsets $\*y_{i,1}$ and $\*y_{i,2}$. 

The subset $\*y_{i,1}$ contains angles index $\{0, 2, 4, \ldots, K-2\}$ and $\*y_{i,2}$ contains angles index ${1, 3, 5, \ldots, K-1}$. The FBP is applied to split noisy sinograms in each subset, yielding two sets of sub-reconstructions
\begin{equation}
	\hat{\*x}_{i,1}=\*R_1 \*y_{i,1}, \quad \hat{\*x}_{i,2}=\*R_2 \*y_{i,2}
\end{equation}

The two sets of sub-reconstructions learn the mapping function from each other, that is, if the batch size equals to 1, each batch in the training contains two input-target pairs $\hat{\*x}_{i,1} \sim \hat{\*x}_{i,2}$ and $\hat{\*x}_{i,2} \sim \hat{\*x}_{i,1}$. 

The optimal weights $\hat{\*\theta}$ of the network $f_{\*\theta}$ are obtained by minimizing the following loss function
\begin{equation}\label{eq:loss_N2I}
	L_{N2I}(\*\theta) = \frac{1}{N}\sum_{i=1}^{N}\| f_{\*\theta}(\hat{\*x}_{i,1}) - \hat{\*x}_{i,2} \|^2_2  +  \| f_{\*\theta}(\hat{\*x}_{i,2}) - \hat{\*x}_{i,1} \|^2_2
\end{equation}

While N2I can effectively remove the noise and find an estimate close to (\ref{eq:FBPnoisy}), it is important to highlight the issue arising when a sparse set of projections $\{\phi_1,\phi_2,\ldots,\phi_K\}$ is used. In this case, $\*x^*$ is a corrupted solution since the operator $\*A$, and therefore $\*R$, are highly ill-conditioned. As a result, the performance of N2I is less robust to generalize to different CT acquisitions in testing. 

In order to improve over noisy artefacts with LDCT, we need to exploit some additional information from the reconstructed image obtained from the CT acquisition geometry; we will introduce the proposed augmentation method based on invariance in the next section.

\begin{figure*}[!h]
	\centering
	\includegraphics[width=\textwidth]{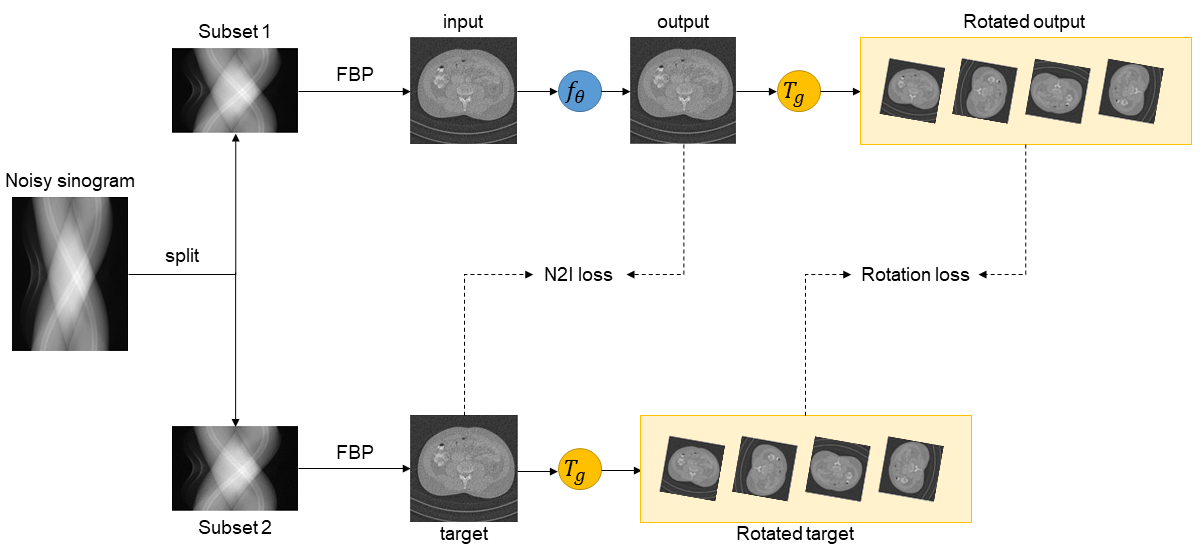}
	\caption{Workflow diagram of the RAN2I framework, considering $r=4$ angular rotations.}\label{fig:EN2I_diagram}
\end{figure*}

\section{Rotational Augmented Noise2Inverse (RAN2I)}\label{ch3_EN2I}

We aim to exploit the property, induced by the CT system, that the images should be invariant to certain groups of transformations such as rotations. A set of images $\-X$ is invariant to the group transformation $g\in\-{G}$ if for all $\*x\in\-X$, $T_g(\*x)\in\-X$ with $T_g$ a unitary matrix. In the noiseless case \cite{cohen2016group,lenc2015understanding}, a function $f$ is equivariant to the transformation $T_g$ if $f(T_g(\*x)) =T_gf(\*x)$ for all $\*x\in \-X$. 

However, enforcing the equivariant property directly into the N2I loss function in the image domain is challenging for the following reasons: 1) the terms of the cost function (\ref{eq:opt_n2i})  measure the error between the output of the network $f_{\theta}$ respect to the noisy input and this makes not possible to embed the equivariance relationship unless employing a second network which would require more computational cost; 2) in the noisy case, with the N2I training model we do not have direct access to the unique image $\*x$ but rather the input and the target are different noisy instances $\hat{\*x}_{i,J^C}$ and $\hat{\*x}_{i,J}$ obtained from the same measurement $\*y$. 

The proposed RAN2I method promotes the equivariance by modifying the N2I training loss function by embedding a term related to rotational augmentation. The proposed term enforces the application of the group transformation $T_g$ to the output of the network $f_{\*\theta}$ for the input image $\hat{\*x}_{i,J^C}$, i.e. $T_g \left(f_{\*\theta}(\hat{\*x}_{i,J^C})\right)$ should be close to the result obtained by applying the same transformation $T_g$ to the target, $T_g(\hat{\*x}_{i,J})$. 

Fig. \ref{fig:EN2I_diagram} shows the diagram of RAN2I for one image set, i.e. $i=1$, where the noisy sinogram is split into $S=2$ subsets by selecting an interleaved set of angles $J$ and $J^C$ and $r=4$ rotations for the transformation $T_g$. The FBP is applied to each sub-sinogram $\*y_{i,J^C}$ and $\*y_{i,J}$ to obtain the noisy reconstructions 
\begin{equation}
	\hat{\*x}_{i,J^C} = \*R_{J^C}\*y_{i,J^C}=\*R_{J^C}\*A_{J^C}\*x_i + \*R_{J^C}\*\epsilon
\end{equation}
where $\*R_{J^C}$ represents the FBP on the angles subset $J^C$, and $\hat{\*x}_{i,J} = \*R_{J}\*y_{i,J}$. 

Finally we construct $T_g\left(f_{\*\theta}(\hat{\*x}_{i,J^C})\right)$ and $T_g(\hat{\*x}_{i,J})$ to impose the equivariance property. In the diagram four rotations $r=4$ transformation is employed, but the framework allows to incorporate different number of rotations or different transformations. 

A simple case to analyse is with large number of angles $K$ and low photons, since 
\begin{equation}	\*R_{J^C}\*A_{J^C}\*x_i\approx\*R_J\*A_J\*x_i \approx\*x_i
\end{equation}
and we enforce $T_g \left(f_{\*\theta}(\*x_i + \*R_{J^C}\*\epsilon)\right)$ to be equal to $T_g(\*x_i + \*R_{J}\*\epsilon)$ meaning that the transformed noisy input should be close to the transformed denoised output of the network $f_{\*\theta}$. 

To promote the equivariance between the rotated output $T_g (f_{\*\theta} (\hat{\*x}_{i,J^C}))$ and the rotated target $T_g (\hat{\*x}_{i,J})$, we add an additional constraint to the N2I loss described in Eq. (\ref{eq:opt_n2i}). Therefore, the optimal network weights are obtained by minimizing the RAN2I training loss:
\begin{align}\label{eq:opt_eqn2i}
	\hat{\*\theta} = \arg\min_{\*\theta}\sum_{i=1}^N\sum_{J\in\-J} & \left(\|f_{\*\theta}(\hat{\*x}_{i,J^C}) - \hat{\*x}_{i,J} \|_2^2 + \right. \\
	& \quad + \left. \|T_g\left(f_{\*\theta}(\hat{\*x}_{i,J^C})\right) - T_g\left(\hat{\*x}_{i,J}\right) \|_2^2\right) \nonumber
\end{align}
The expected prediction error can be decomposed into two parts according to the following proposition.

\begin{prop}[Expected  prediction  error  decomposition] \label{prop:pred_error}
	Let $\hat{\*x}_{J} = \*R_{J}\*y_{J}$, $\hat{\*x}_{J^C} = \*R_{J^C}\*y_{J^C}$, ${\*x}^* = \*R\*y^*$, $\*R_J$ be a linear operator $\forall J$ and $\bm\epsilon$ be element-wise independent, then we have
	\begin{align}\label{eq:pred_error}
		\@E_{x,\epsilon} \| f_{\*\theta}(\hat{\*x}_{J^C}) - \hat{\*x}_{J} \|^2 + \@E_{x,\epsilon} \| T_g f_{\*\theta}(\hat{\*x}_{J^C}) - T_g (\hat{\*x}_{J}) \|^2 \nonumber \\ 
		= \@E_{x,\epsilon} \|  f_{\*\theta}(\hat{\*x}_{J^C}) - {\*x}^*_J \|^2 + \@E_{x,\epsilon} \| \*x^*_J - \hat{\*x}_{J} \|^2 \nonumber \\
		+ \, \@E_{x,\epsilon} \| T_g f_{\*\theta}(\hat{\*x}_{J^C}) - T_g ({\*x}^*_J) \|^2 \nonumber \\ 
		+ \@E_{x,\epsilon} \| T_g ({\*x}^*_J) - T_g (\hat{\*x}_{J}) \|^2
	\end{align}
\end{prop}
 
\begin{proof}
	The proof is detailed in Appendix \ref{app_conv}.
\end{proof}

Proposition \ref{prop:pred_error}  states that the expected overall prediction error can be decomposed into the weighted sum of the supervised prediction error with respect to $\*x^*_J$ for the original and rotational transformed image. These terms depend on the choice of $f_{\*\theta}$ while the variance of the reconstruction noise does not depend on $f_{\*\theta}$. Therefore, when minimizing (\ref{eq:pred_error}), the function $f_{\*\theta}$ minimizes the difference between its output and the unknown target sub-reconstruction. The equivariant term aims to reduce the prediction error respect to the true $\*x$, which especially in the sparse view case lie in a solution space far from the target sub-reconstruction $\*x^*_J$. 

\section{Setups for simulation and experiment} \label{sec:Setups}

\subsection{Training}

All the simulations were carried out in Python 3.8 with an NVIDIA GeForce RTX 3070 GPU. For the neural network $f_{\*\varphi}$ we employed the Bias-free DnCNN (BF-DnCNN) architecture \cite{mohan2019robust} which promotes better generalization. 

\begin{figure}[!h]
    \centering
    \begin{minipage}{.5\textwidth}
    \includegraphics[scale=0.45]{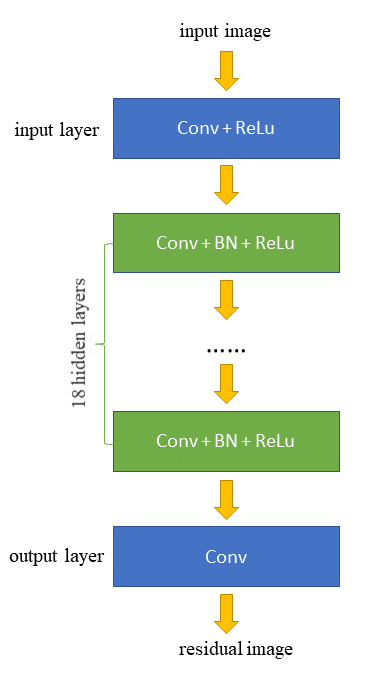}
    \includegraphics[scale=0.45]{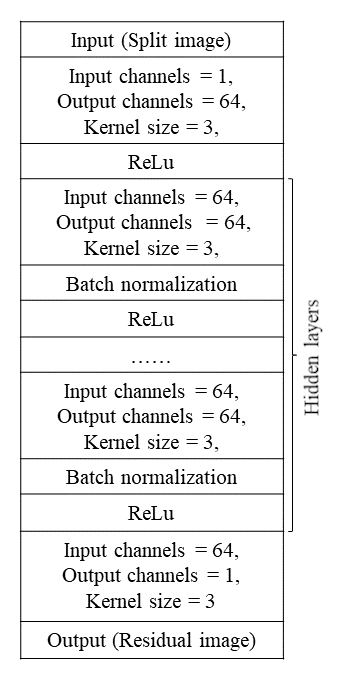}
    \caption{Summary of DnCNN architecture for RAN2I.}
    \label{fig:architecture}
    \end{minipage}
\end{figure}

The BF-DnCNN model consists of 20 convolutional layers, each one constituted of 64 filters of size $3\times 3$, batch normalization \cite{ioffe2015batch} and ReLU function, but without any additive bias terms, including the batch normalization of each layer. Fig. \ref{fig:architecture} shows the architecture.

The training was performed in Pytorch using Adam optimizer \cite{kingma2014adam} with the learning rate of $10^{-3}$ and batch size equal to 1. The network was trained for 100 epochs in total; the mean squared error was used to compute the loss. In all the simulations for RAN2I, we fixed the number of splits $S=2$ to generate the partitioned sinograms, hence the only combination strategy for sub-reconstructions is "1:1".

The above described common settings have been used for training and validation of numerical simulations and real data experiments. Section \ref{sec: numerical results} shows the results of numerical simulation and section \ref{sec: experimental results} shows the results of the experiment using real data acquisition. \footnote{Our code is publicly available at \url{https://github.com/UoD-MCI/RAN2I}.}

\subsection{Implementation of rotation}

In the implementation of fixed or random group rotations, the number of rotations $r$ are hyper-parameters. Each random rotation ranges from $1$ to $360$ degrees; each fixed rotation ranges from $30$ to $360$ degrees and the interval is $\frac{360}{r}$ degrees where $r$ represents the number of rotations. For the hyper-parameters of RAN2I in the simulation, we used the random rotation and the number of rotations $r=2$ as default settings.

\section{Numerical simulation results} \label{sec: numerical results}

\subsection{Data processing}

We performed our numerical simulations using the full-dose CT images obtained from the 2016 NIH-AAPM-Mayo Clinic Low Dose CT Grand Challenge \cite{moen2021low} which contains 10 patients, and 2378 slices of CT images in total. We used 916 slices from 4 patients as the training dataset and one additional patient as the testing dataset. We used a resolution of $512 \times 512$ in the image domain. 

\begin{figure}[!h]
    \centering
    \begin{minipage}{.5\textwidth}
    \includegraphics[width=\textwidth]{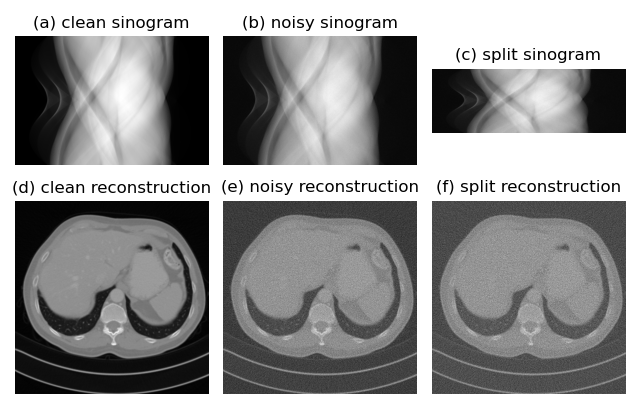}
    \caption{Examples of split sinogram and reconstruction. a) The simulated clean sinogram obtained from the full-dose image when sampling angles $K=512$. b) The noisy sinogram corrupted by Poisson noise when $I_0=10^4$. c) Split noisy sinogram when $S=2$. d-f) reconstructions of sinograms a-c) respectively.}
    \label{fig:split_example}
    \end{minipage}
\end{figure}

The sinograms were obtained from the above images using a 2D parallel-beam geometry which was implemented with the ASTRA Toolbox \cite{van2015astra}. The simulated detector array consists of 768 detectors with a width of 1 mm each. The sampling angles and photon counts are set depending on different experiments such as Poisson noise and sparse-view CT. After obtaining the noisy sinograms, the sinograms were split into two subsets as described in section \ref{subsec:N2I} and reconstructed using FBP to obtain two sets of sub-reconstructions to be used as the dataset for training the neural network. The testing dataset was processed the same way as the training dataset. Fig. \ref{fig:split_example} shows examples of split data.

\begin{figure*}[!h]
	\begin{center}
	\small\addtolength{\tabcolsep}{-18pt}
	\renewcommand{\arraystretch}{0.1}
	
	\begin{tabular}{cccccc}
		\hspace{-.4cm}\small{(a) Ground Truth} 
		& 
		\specialcell[c]{\hspace{-.6cm}\small (b) FBP}
		& 
		\specialcell[c]{\hspace{-.6cm}\small (c) N2I} 
		& 
		\specialcell[c]{\hspace{-.7cm}\small (d) Score Diffusion}   
        & 
		\specialcell[c]{\hspace{-.7cm}\small (e) FBPConvNet}   
        & 
		\specialcell[c]{\hspace{-.7cm}\small (f) RAN2I} \\
		\vspace{.2cm}
		\hspace{.4cm} \begin{tikzpicture}  
				\node {\includegraphics[width=0.16\textwidth]{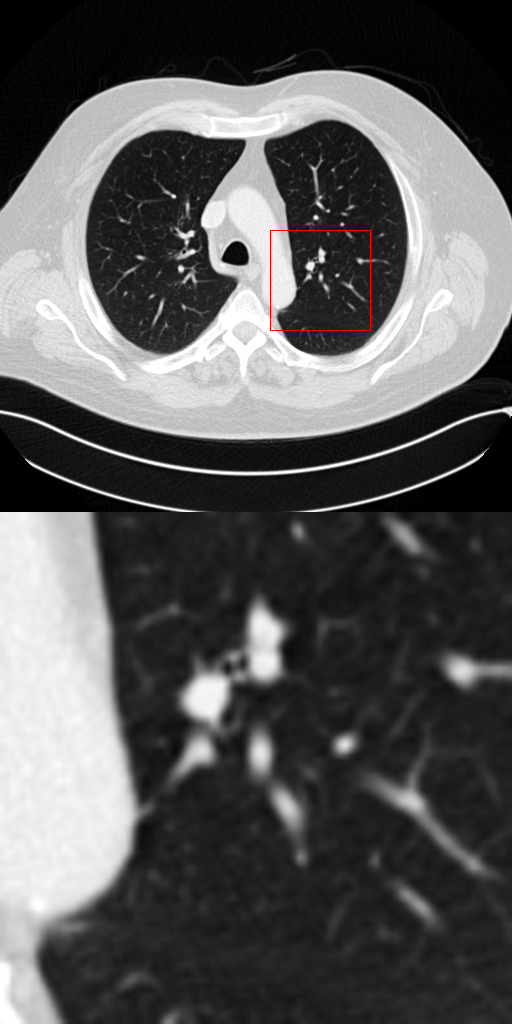}};
		\end{tikzpicture} \hspace{.55cm}
		&
		\begin{tikzpicture}
			\node {\includegraphics[ width=0.16\textwidth]{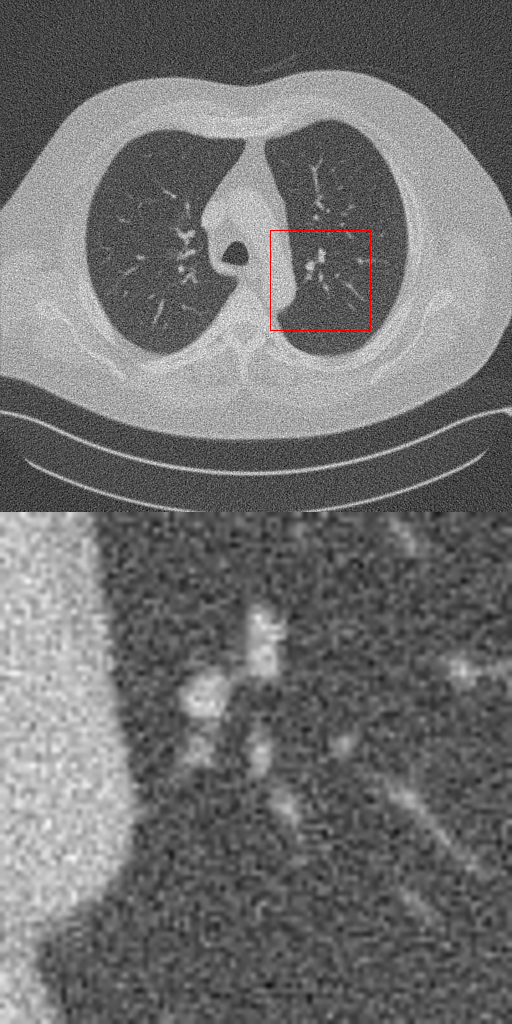}};
		\end{tikzpicture} \hspace{.55cm}
		&
		\begin{tikzpicture}
			\node {\includegraphics[width=0.16\textwidth]{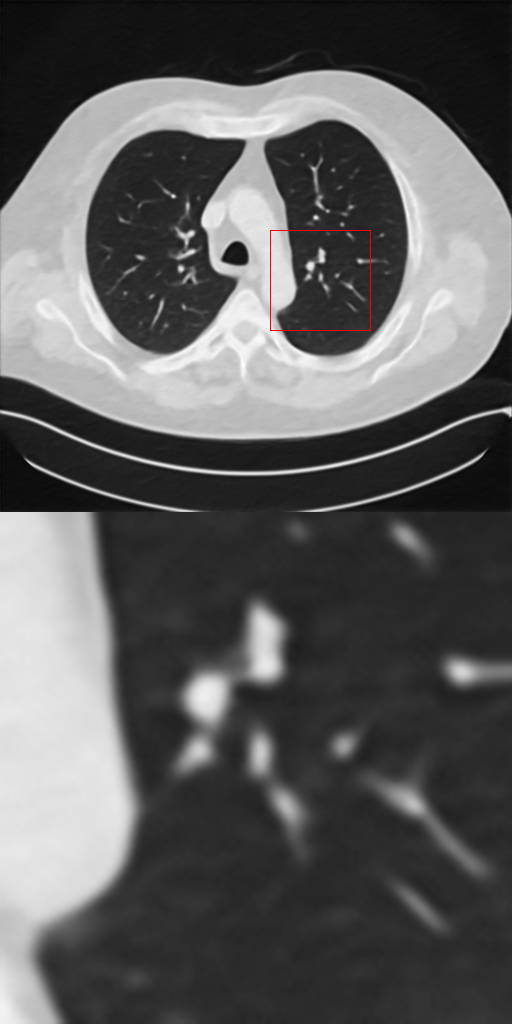}};
		\end{tikzpicture} \hspace{.55cm}
		&       
        \begin{tikzpicture}  
				\node {\includegraphics[width=0.16\textwidth]{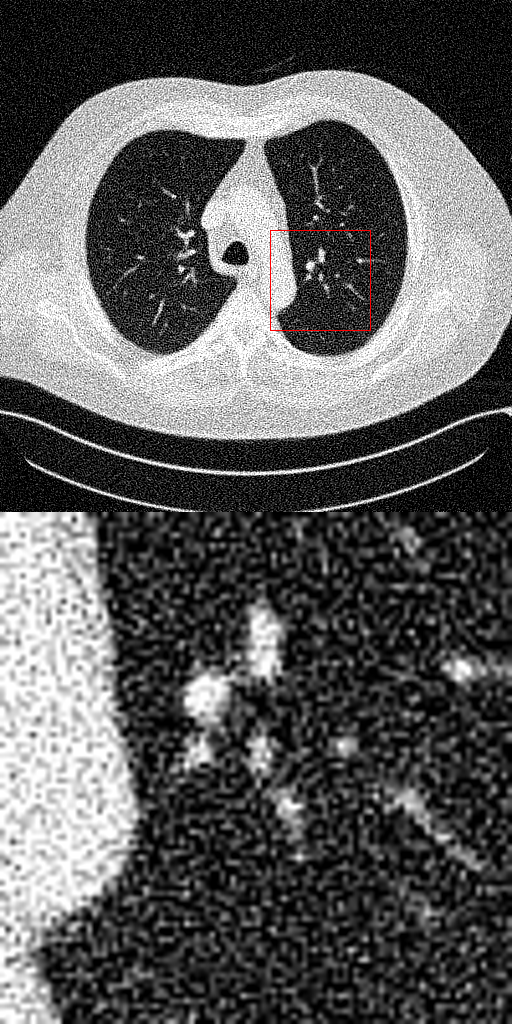}};
		\end{tikzpicture} \hspace{.55cm}
        &       
        \begin{tikzpicture}
			\node {\includegraphics[width=0.16\textwidth]{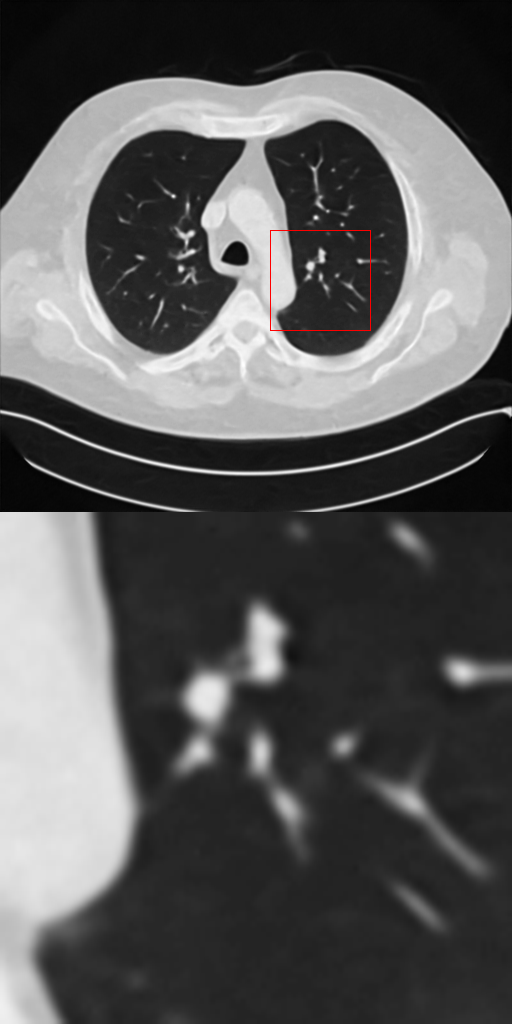}};
		\end{tikzpicture} \hspace{.55cm}
        &       
        \begin{tikzpicture}  
				\node {\includegraphics[width=0.16\textwidth]{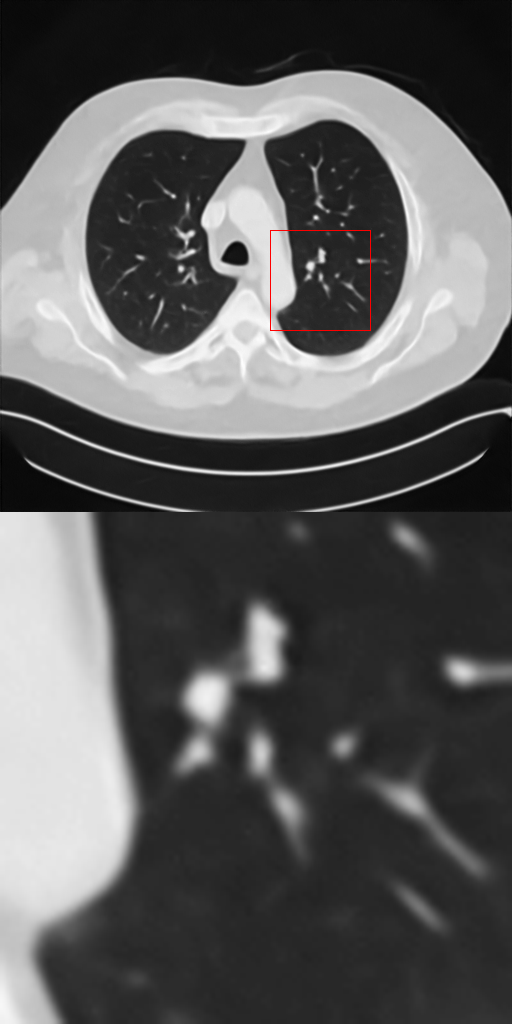}}; 
		\end{tikzpicture} \hspace{.55cm}
		\\   		
	\end{tabular}
	\caption{Qualitative CT reconstruction results with low photon count Poisson noise: a) ground truth, b) FBP, c) N2I, d) Score-based diffusion model, e) FBPConvNet and f) RAN2I. The Hounsfield unit (HU) display range is set to [-1900, 900].} \label{fig:result_poisson}
 \end{center}
\end{figure*}

\subsection{Comparison between CT reconstruction methods}

The proposed RAN2I algorithm was compared with FBP, N2I, the supervised approach FBPConvNet \cite{jin2017deep}, and the score-based diffusion model for inverse problems \cite{song2021solving}.

\subsubsection{Filtered back-projection}

The FBP for comparison was applied directly on complete noisy sinograms using ASTRA Toolbox \cite{van2015astra} whose geometry settings are introduced in section \ref{sec: numerical results}. A. We used a Ram-Lak filter for FBP.

\subsubsection{Noise2Inverse}

N2I for comparison also used $S=2$ splits to partition noisy sinograms. The training setups of N2I were exactly the same as RAN2I but the loss function is only the mean square error (MSE) as described in Eq. (\ref{eq:loss_N2I}) without the rotational augmented term. Therefore, the comparison between N2I and RAN2I in section \ref{sec: numerical results} can be regarded as the ablation study on rotational augmentation.

\subsubsection{Supervised Learning}

For comparison with the supervised method, we implemented FBPConvNet strategy and created another dataset in which the input and target are noisy and clean reconstructions respectively. The noisy images are generated using FBP with low photon counts or sparse angles depending on the simulations. 
The clean images are generated as the target using FBP with $K=1024$ angles but without any noise. The training configurations, i.e. network architecture, optimizer, batch size, and training epochs are exactly the same as RAN2I except for the loss function using MSE without the rotational augmented term.

\subsubsection{Score-based diffusion model}

The proposed RAN2I was also compared with the state-of-the-art score-based diffusion model for inverse problem \cite{song2021solving}. We did not train a new model as their pre-trained model is based on the same dataset we used. We directly used their officially released codes in our simulations. For sparse-view simulation, we only need to change the sampling angles to reproduce the result. For Poisson noise simulation, we modified the codes to simulate Poisson noise in the raw data acquisition as they did not consider Poisson noise in their experiments.

\subsection{Results}

Peak signal-to-noise ratio (PSNR) and structural similarity index (SSIM), calculated with reference to the full test image, were used to evaluate the quality of the results. 
We selected 50 slices from the test patient as testing datasets which was processed the same way as training datasets and computed the mean PSNR and SSIM of outputs from the network.

\subsubsection{Poisson noise}

To evaluate the ability to remove Poisson noise, we simulated low-quality images using low photon counts $I_0=10^4$ and high angles $K=1024$ in both training and testing. Fig. \ref{fig:result_poisson} shows the results of different methods. Using FBP directly will result in heavy noise on the CT image. The result of score-based model is still corrupted by noise as score-based model only considers Gaussian noise or is used in image restoration tasks such as sparse-view CT. Both N2I and RAN2I remove the noise but our proposed RAN2I produces a clearer image with better contrast. Also, the image quality of RAN2I is comparable to supervised method. 
Table \ref{tab:poisson} shows the PSNR and SSIM of different methods. Our proposed RAN2I method has the highest SSIM and the PSNR is comparable to supervised method and better than other methods.

\begin{table}[!h]
	\caption{Quantitative results with $I_0=10^4$ and $K=1024$ with parallel beam CT geometry.}\label{tab:poisson}
	\begin{minipage}{.5\textwidth}
        \begin{center}
     	\setlength{\tabcolsep}{10pt} 
		  \renewcommand{\arraystretch}{1.5} 
		  {\begin{tabular}{l|ll}
                \hline
                Methods     & PSNR & SSIM \\ \hline \hline
                FBP         & $18.05\pm 0.67$     & $0.32\pm 0.01$     \\
                N2I         & $37.97\pm 0.95$     & $0.93\pm 0.01$     \\
                Score-based & $17.80\pm 0.53$     & $0.43\pm 0.03$     \\
                FBPConvNet (Supervised) & $39.00\pm 0.93$     & $0.94\pm 0.01$
                \\
                \textbf{RAN2I}       & $\textbf{38.37}\pm 1.06$     & $\textbf{0.94}\pm 0.01$     \\ \hline
            \end{tabular}}
        \end{center}
	\end{minipage}
\end{table}

\begin{figure*}[!h]
	\begin{center}
	\small\addtolength{\tabcolsep}{-18pt}
	\renewcommand{\arraystretch}{0.1}
	
	\begin{tabular}{cccccc}
		\hspace{-.4cm}\small{(a) Ground Truth} 
		& 
		\specialcell[c]{\hspace{-.6cm}\small (b) FBP}
		& 
		\specialcell[c]{\hspace{-.6cm}\small (c) N2I} 
		& 
		\specialcell[c]{\hspace{-.7cm}\small (d) Score Diffusion}   
        & 
		\specialcell[c]{\hspace{-.7cm}\small (e) FBPConvNet}   
        & 
		\specialcell[c]{\hspace{-.7cm}\small (f) RAN2I} \\
		\vspace{.2cm}
		\hspace{.4cm} \begin{tikzpicture}  
				\node {\includegraphics[width=0.16\textwidth]{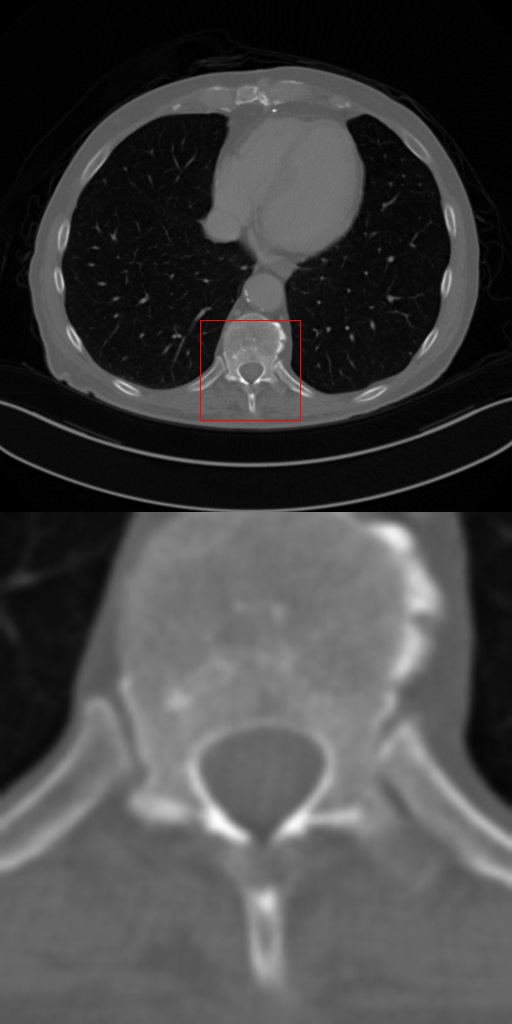}};
		\end{tikzpicture} \hspace{.55cm}
		&
		\begin{tikzpicture}
			\node {\includegraphics[ width=0.16\textwidth]{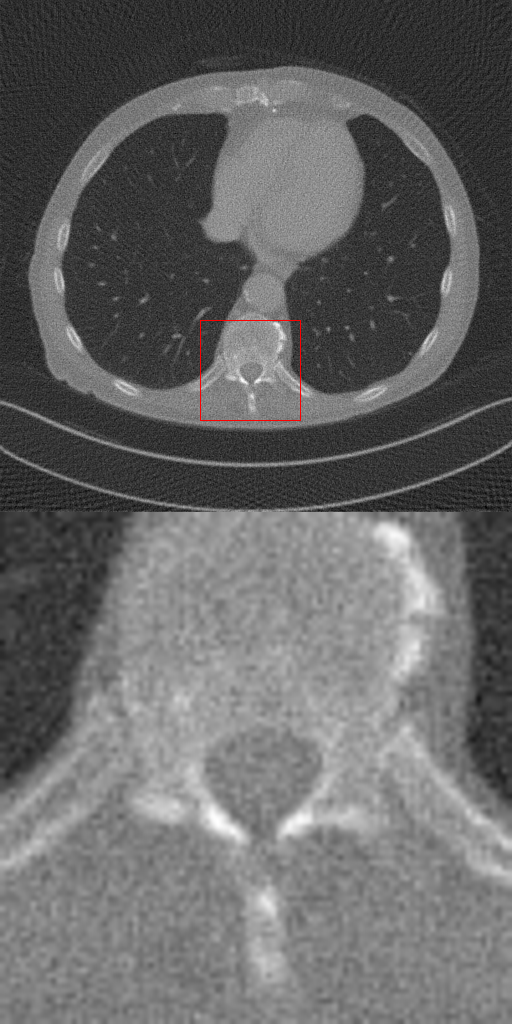}};
		\end{tikzpicture} \hspace{.55cm}
		&
		\begin{tikzpicture}
			\node {\includegraphics[width=0.16\textwidth]{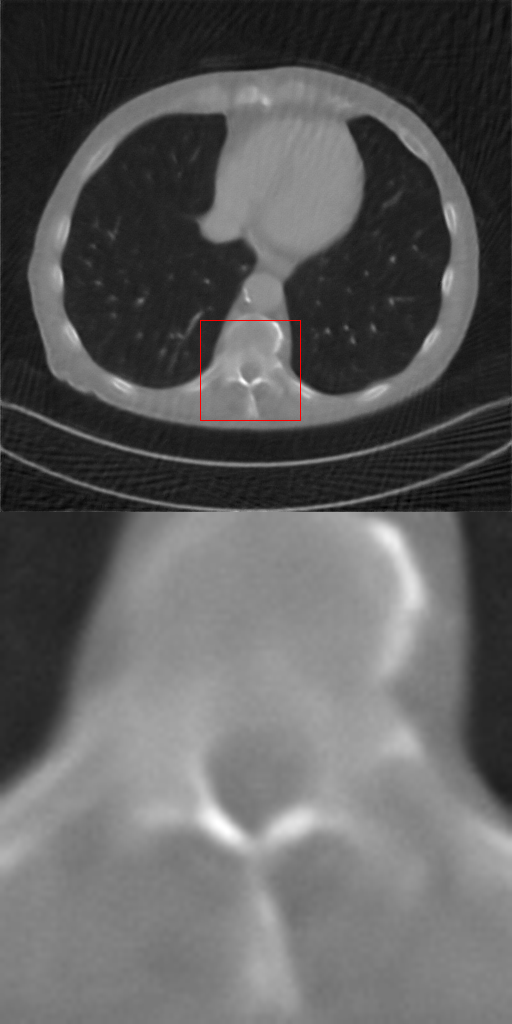}};
		\end{tikzpicture} \hspace{.55cm}
		&       
        \begin{tikzpicture}  
				\node {\includegraphics[width=0.16\textwidth]{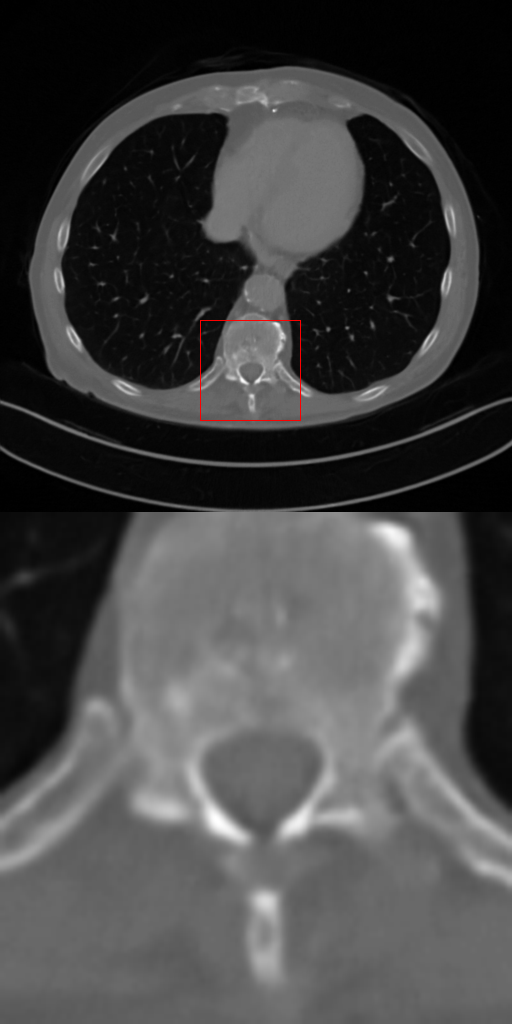}};
		\end{tikzpicture} \hspace{.55cm}
        &       
        \begin{tikzpicture}
			\node {\includegraphics[width=0.16\textwidth]{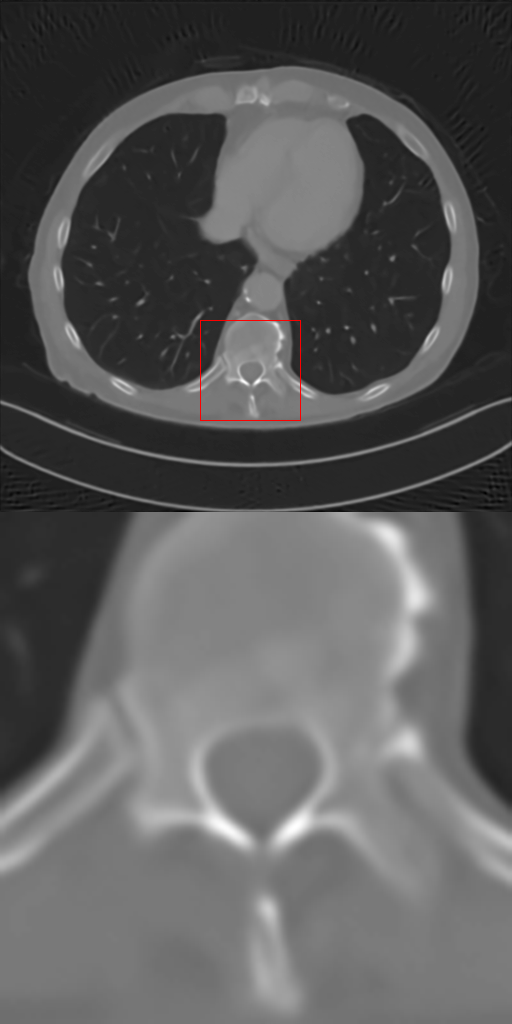}};
		\end{tikzpicture} \hspace{.55cm}
        &       
        \begin{tikzpicture}  
				\node {\includegraphics[width=0.16\textwidth]{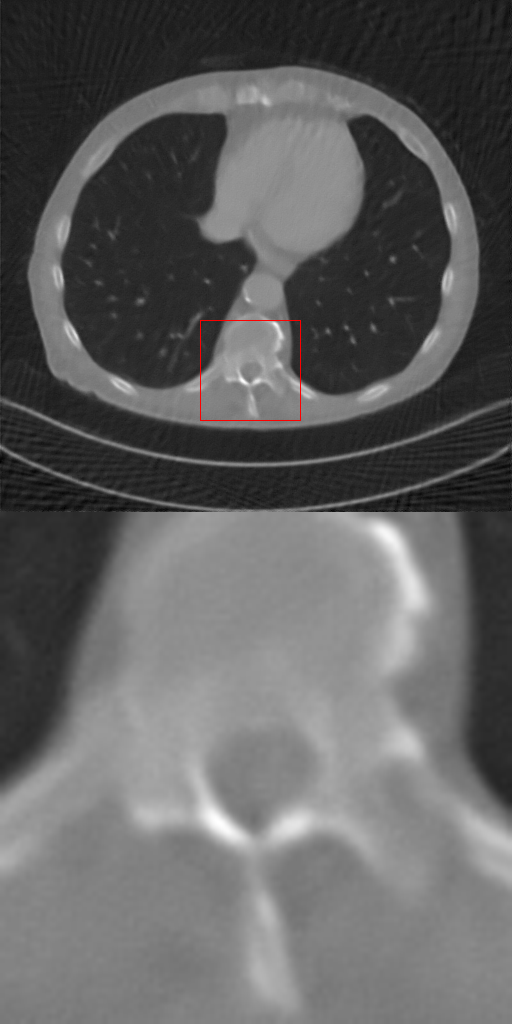}}; 
		\end{tikzpicture} \hspace{.55cm}
		\\   		
	\end{tabular}
	\caption{Qualitative results for sparse-view CT reconstruction: a) ground truth, b) FBP, c) N2I, d) Score-based Diffusion model, e) FBPConvNet and f) RAN2I. The HU display range is set to [-1991, 1875].} \label{fig:result_sparseview}
 \end{center}
\end{figure*}

\begin{table*}[!b]
        \renewcommand\thetable{III}
	\caption{Quantitative results with $I_0=10^4$ and different number of angles with parallel beam CT geometry.}\label{tab:res1}
	\begin{center}
		\setlength{\tabcolsep}{5pt} 
		\renewcommand{\arraystretch}{1.1} 
		\begin{tabular}{c|cccc||cccc}
			\hline \hline
			& \multicolumn{4}{c||}{Training: $K=512$ angles, $I_0 = 10^4$  photons} & \multicolumn{4}{c}{Training: $K=1024$ angles, $I_0 = 10^4$  photons} \\ \hline \hline
			angles & 128 & 256 & 512 & 1024 & 128 & 256 & 512 & 1024 \\ \hline
			& \multicolumn{4}{c||}{PSNR} & \multicolumn{4}{c}{PSNR} \\ \hline
			RAN2I & $\mathbf{31.66}\pm 0.69$ & $\mathbf{35.18}\pm 0.86$ & $\mathbf{36.70}\pm 0.99$ & $\mathbf{37.21}\pm 1.01$ & $\mathbf{31.13}\pm 0.70$ & $\mathbf{34.63}\pm 0.73$ & $\mathbf{37.16}\pm 0.97$ & $\mathbf{38.37}\pm 1.06$\\
			N2I & $30.56\pm 0.61$ & $33.85\pm 0.74$ & $35.83\pm 0.86$ & $36.88\pm 0.94$ & $30.14\pm 0.60$ & $34.09\pm 0.67$ & $36.52\pm 0.83$ & $37.97\pm 0.95$  \\
            Sup & $34.67\pm 0.71$ & $36.72\pm 0.84$ & $37.98\pm 0.96$ & $38.58\pm 1.03$ & $33.23\pm 0.54$ & $35.66\pm 0.61$ & $37.66\pm 0.75$ & $39.00\pm 0.93$ \\ \hline
			& \multicolumn{4}{c||}{SSIM} & \multicolumn{4}{c}{SSIM}\\ \hline
			RAN2I & $\mathbf{0.80}\pm 0.02$ & $\mathbf{0.89}\pm 0.01$ & $\mathbf{0.92}\pm 0.01$ & $\mathbf{0.94}\pm 0.01$ & $\mathbf{0.81}\pm 0.02$ & $\mathbf{0.88}\pm 0.01$ & $\mathbf{0.92}\pm 0.01$ & $\mathbf{0.94}\pm 0.01$ \\
			N2I & $0.75\pm 0.02$ & $0.87\pm 0.01$ & $0.91\pm 0.01$ & $0.93\pm 0.01$ & $0.75\pm 0.01$ & $0.86\pm 0.01$ & $0.91\pm 0.01$ & $0.93\pm 0.01$   \\
            Sup & $0.87\pm 0.01$ & $0.92\pm 0.01$ & $0.93\pm 0.01$ & $0.94\pm 0.01$ & $0.83\pm 0.01$ & $0.89\pm 0.01$ & $0.93\pm 0.01$ & $0.94\pm 0.01$ \\
			\hline \hline 
		\end{tabular}
	\end{center}
\end{table*}

\begin{table}[!h]
        \renewcommand\thetable{II}
	\caption{Quantitative results with $I_0=10^5$ and $K=128$ with parallel beam CT geometry.}\label{tab:sparseview}
	\begin{minipage}{.5\textwidth}
        \begin{center}
     	\setlength{\tabcolsep}{10pt} 
		  \renewcommand{\arraystretch}{1.5} 
		  {\begin{tabular}{l|ll}
                \hline
                Methods     & PSNR & SSIM \\ \hline \hline
                FBP         & $20.34\pm 0.37$     & $0.69\pm 0.01$     \\
                N2I         & $22.70\pm 1.53$     & $0.86\pm 0.04$     \\
                Score-based & $35.29\pm 0.60$     & $0.95\pm 0.01$     \\
                FBPConvNet (Supervised)  & $27.77\pm 0.60$     & $0.92\pm 0.02$
                \\
                \textbf{RAN2I}      & $\textbf{22.86}\pm 1.05$     & $\textbf{0.86}\pm 0.03$   \\ \hline
            \end{tabular}}
        \end{center}
	\end{minipage}
\end{table}

\subsubsection{Sparse-view CT}

In the sparse-view CT experiment, the low-quality images were obtained using limited angles $K=128$ and normal photon counts $I_0=10^5$. Fig. \ref{fig:result_sparseview} shows that the score-based diffusion model has state-of-the-art performance in the sparse-view CT. The supervised method also removes the artifacts but the result is smooth. N2I has the most blurry result in sparse-view CT. RAN2I has improvement over N2I, as it can be observed that some details have a more precise edge. Table \ref{tab:sparseview} shows the quantitative results in the sparse-view CT experiment where the score-based method demonstrates superior results and RAN2I still has improvement over N2I in sparse-view CT.

\subsection{Generalization study}

\subsubsection{Test with different angles and photon counts}

To show the improvement of our method RAN2I with respect to N2I and how the results generalize to different physical CT settings in testing, we trained the network with datasets obtained from $K=512,1024$ angles and $I_0=10^4$ photon counts, then we tested the methods with different sampling angles and photon counts $I_0$.

Table \ref{tab:res1} shows the mean PSNR and SSIM with the standard deviation of RAN2I and N2I when testing with fixed photon counts $I_0=10^4$  and different angles acquisition, i.e. mismatched CT acquisition system. Numbers in bold are results of the proposed RAN2I method.

When the network is trained with $K=1024$ angles, RAN2I shows comparable performances to N2I. 
However when lowering the number of angles in the training dataset $K=512$, RAN2I noticeably shows a PSNR improvement with respect to N2I, consistently for different CT settings, and the improvement is increasing significantly for fewer views, $K\leq 256$. 
These results demonstrate our intuition that the equivariance property in RAN2I improves the accuracy, especially with a reduced number of measurements.

\begin{table*}[!h]
	\caption{Quantitative results with the same angles and different amount of photon counts with parallel beam CT geometry.}\label{tab:res2}
	\begin{center}
		\setlength{\tabcolsep}{5pt} 
		\renewcommand{\arraystretch}{1.1} 
		\begin{tabular}{c|cccc||cccc}
			\hline \hline
			& \multicolumn{4}{c||}{Training: $K=512$ angles, $I_0 = 10^4$  photons} & \multicolumn{4}{c}{Training: $K=1024$ angles, $I_0 = 10^4$  photons} \\ \hline \hline
			$I_0$ & $10^3$ & $5\cdot 10^3$ & $10^4$ & $10^5$ & $10^3$ & $5\cdot 10^3$ & $10^4$ & $10^5$ \\ \hline 	 
			& \multicolumn{4}{c||}{PSNR} & \multicolumn{4}{c}{PSNR} \\ \hline
			RAN2I & $\mathbf{30.68}\pm 0.63$ & $\mathbf{35.45}\pm 0.89$ & $\mathbf{36.70}\pm 0.98$ & $\mathbf{37.65}\pm 1.02$ & $\mathbf{32.67}\pm 0.67$ & $\mathbf{37.32}\pm 0.97$ & $\mathbf{38.37}\pm 1.06$ & $\mathbf{39.50}\pm 1.07$ \\
			N2I & $29.69\pm 0.52$ & $34.65\pm 0.78$ & $35.83\pm 0.85$ & $36.86\pm 0.95$ & $31.74\pm 0.51$ & $36.72\pm 0.85$ & $37.97\pm 0.95$ & $39.05\pm 1.01$ \\
            Sup & $31.62\pm 0.50$ & $36.71\pm 0.86$ & $37.98\pm 0.96$ & $39.50\pm 1.03$ & $32.60\pm 0.47$ & $37.64\pm 0.75$ & $39.01\pm 0.91$ & $40.85\pm 1.07$ \\
			\hline 
			& \multicolumn{4}{c||}{SSIM} & \multicolumn{4}{c}{SSIM} \\ \hline
			RAN2I & $\mathbf{0.74}\pm 0.01$ & $\mathbf{0.90}\pm 0.01$ & $\mathbf{0.92}\pm 0.01$ & $\mathbf{0.95}\pm 0.01$ & $\mathbf{0.82}\pm 0.01$ & $\mathbf{0.92}\pm 0.01$ & $\mathbf{0.94}\pm 0.01$ & $\mathbf{0.96}\pm 0.01$ \\
			N2I & $0.66\pm 0.02$ & $0.88\pm 0.01$ & $0.91\pm 0.01$ & $0.95\pm 0.01$ & $0.76\pm 0.01$ & $0.91\pm 0.01$ & $0.93\pm 0.01$ & $0.95\pm 0.01$ \\
            Sup & $0.78\pm 0.01$ & $0.92\pm 0.01$ & $0.93\pm 0.01$ & $0.96\pm 0.01$ & $0.81\pm 0.01$ & $0.93\pm 0.01$ & $0.94\pm 0.01$ & $0.97\pm 0.01$ \\
            \hline \hline
		\end{tabular}
	\end{center}
\end{table*}

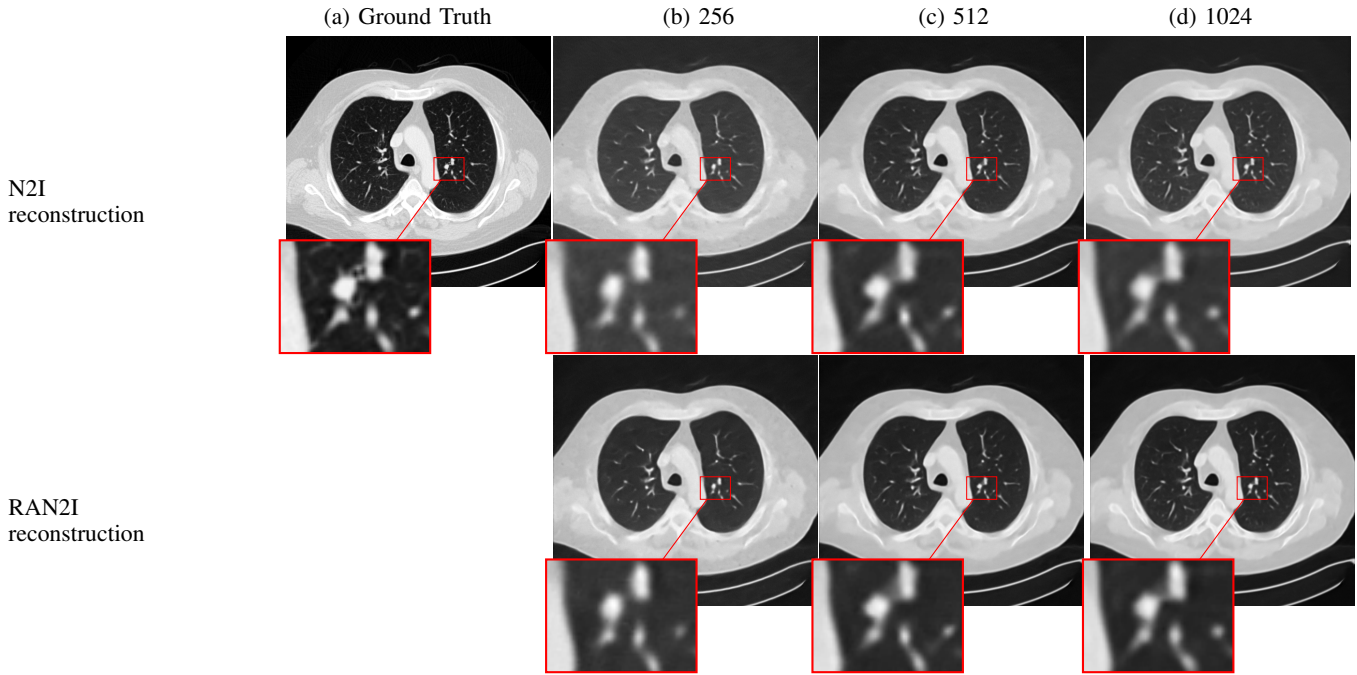
\begin{figure*}[!h]
	\centering
	\small\addtolength{\tabcolsep}{-18pt}
	\renewcommand{\arraystretch}{0.1}
	
	\begin{tabular}{clccc}
		& 
		\specialcell[c]{\hspace{.6cm}\small(a) Ground Truth} 
		&
		\specialcell[c]{\hspace{-.3cm}\small (b) 256}
		& 
		\specialcell[c]{\hspace{-.6cm}\small (c) 512} 
		& 
		\specialcell[c]{\hspace{-.6cm}\small (d) 1024} 
        \\
		\vspace{-.2cm}
		\begin{tikzpicture}
			\node[text width = 0.25\textwidth, text depth = 2cm, anchor=north] {\small\hspace{.1cm}   N2I\\\hspace{.1cm}  reconstruction};
		\end{tikzpicture} 
	    &
		\begin{tikzpicture}
			\begin{scope}[spy using outlines={rectangle,red,magnification=5,width=2cm, height=1.5cm,connect spies}]   
				\node {\includegraphics[viewport=0 10 400 360, clip, width=0.21\textwidth]{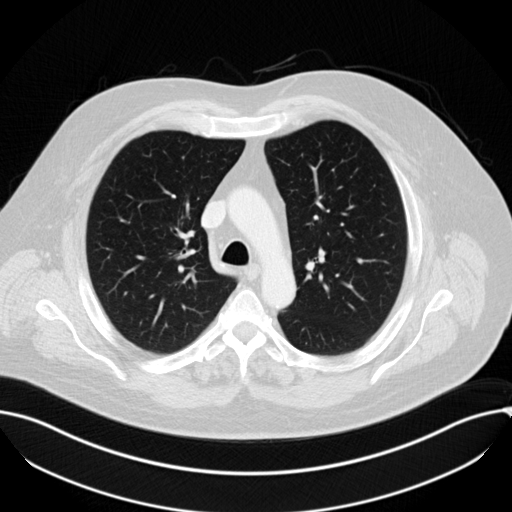}}; 
				\spy on (.25,-.1) in node [left] at (-0.0,-1.8); 
			\end{scope}
		\end{tikzpicture} \hspace{.25cm}
	    &
		\begin{tikzpicture}
			\begin{scope}[spy using outlines={rectangle,red,magnification=5,width=2cm, height=1.5cm,connect spies}]   
				\node {\includegraphics[viewport=0 10 400 360, clip, width=0.21\textwidth]{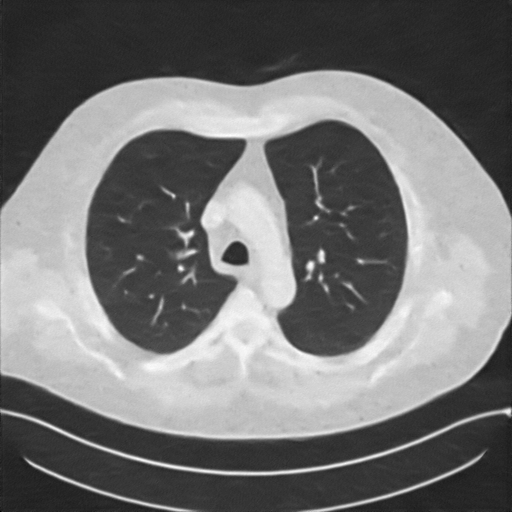}};
				\spy on (.25,-.1) in node [left] at (-0.0,-1.8); 
			\end{scope}
		\end{tikzpicture} \hspace{.25cm}
		&
		\begin{tikzpicture}
			\begin{scope}[spy using outlines={rectangle,red,magnification=5,width=2cm, height=1.5cm,connect spies}]   
				\node {\includegraphics[viewport=0 10 400 360, clip, width=0.21\textwidth]{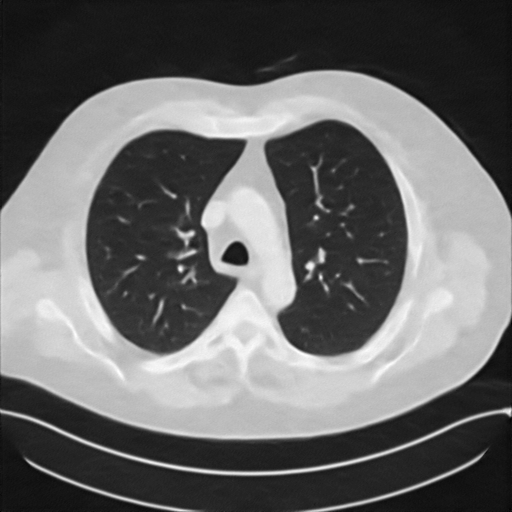}};
				\spy on (.25,-.1) in node [left] at (-0.0,-1.8); 
			\end{scope}
		\end{tikzpicture} \hspace{.25cm}
		&
		\begin{tikzpicture}
			\begin{scope}[spy using outlines={rectangle,red,magnification=5,width=2cm, height=1.5cm,connect spies}]   
				\node {\includegraphics[viewport=0 10 400 360, clip, width=0.21\textwidth]{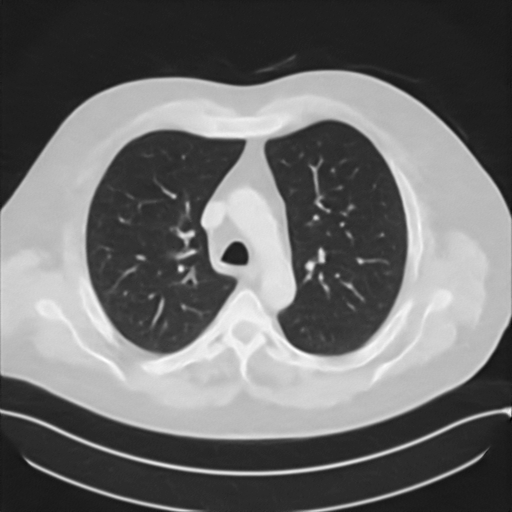}};
				\spy on (.25,-.1) in node [left] at (-0.0,-1.8); 
			\end{scope}
		\end{tikzpicture} 
		\vspace{.1cm} 
		\\ 
		\begin{tikzpicture}
			\node[text width = 0.25\textwidth, text depth = 2cm, anchor=north] {\small\hspace{.1cm}   RAN2I\\\hspace{.1cm}  reconstruction};
		\end{tikzpicture} 
	    &
		\begin{tikzpicture}
			\node[text centered,text width = 0.15\textwidth, text depth = 2cm, anchor=north] {};
		\end{tikzpicture} 
        &
		\begin{tikzpicture}
			\begin{scope}[spy using outlines={rectangle,red,magnification=5,width=2cm, height=1.5cm,connect spies}]   
				\node {\includegraphics[viewport=0 10 400 360, clip, width=0.21\textwidth]{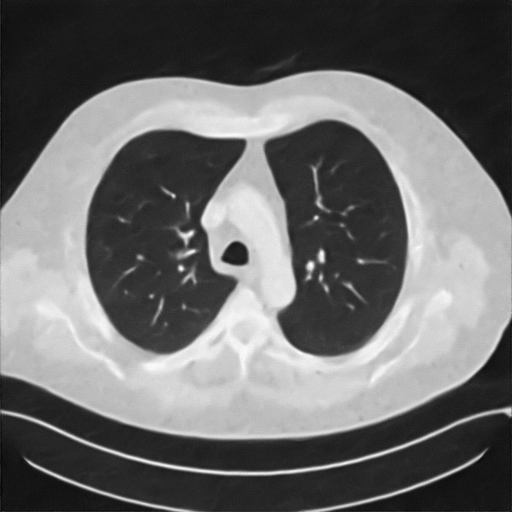}};
				\spy on (.25,-.1) in node [left] at (-0.0,-1.8); 
			\end{scope}
		\end{tikzpicture} \hspace{.25cm}
		&
		\begin{tikzpicture}
			\begin{scope}[spy using outlines={rectangle,red,magnification=5,width=2cm, height=1.5cm,connect spies}]   
				\node {\includegraphics[viewport=0 10 400 360, clip, width=0.21\textwidth]{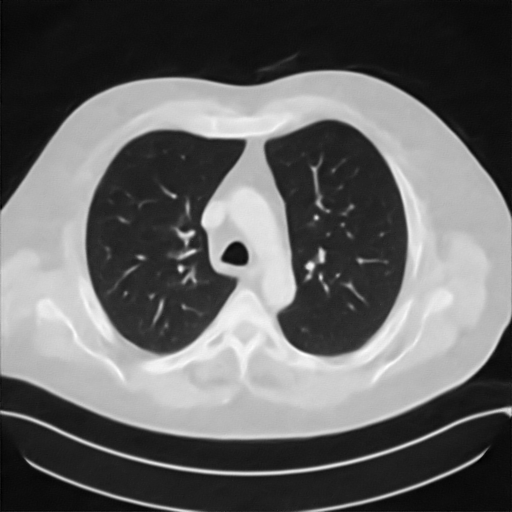}};
				\spy on (.25,-.1) in node [left] at (-0.0,-1.8); 
			\end{scope}
		\end{tikzpicture} \hspace{.25cm}  
		&       
		\begin{tikzpicture}
			\begin{scope}[spy using outlines={rectangle,red,magnification=5,width=2cm, height=1.5cm,connect spies}]   
				\node {\includegraphics[viewport=0 10 400 360, clip, width=0.21\textwidth]{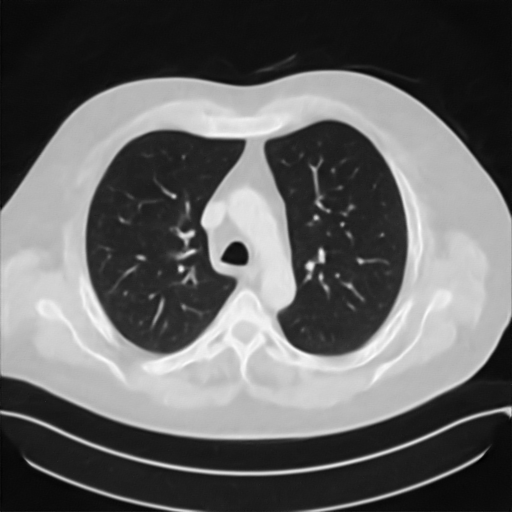}};
				\spy on (.25,-.1) in node [left] at (-0.0,-1.8); 
			\end{scope}
		\end{tikzpicture} 
		
	\end{tabular}
	\caption{Generalization analysis: qualitative reconstruction using $I_0=10^4$ photons counts and $K=512$ angles for different numbers of angles considered in testing (mismatched settings). a) represents the ground truth image, b-d) reconstructions using 256, 512, 1024 projections using N2I (top row) and RAN2I (bottom row). The HU display range is [-1900, 900].} \label{fig:results_diff_proj}
\end{figure*}

Table \ref{tab:res2} shows the results with the same angles in training and testing, $K=512$ and different photon counts $I_0$ in the testing. It is important to note how RAN2I consistently improves over N2I and manifests good generalization capabilities in a wide region of $I_0$.

Fig. \ref{fig:results_diff_proj} shows an example of the generalization behavior of the proposed method under different sampling angles (mismatched settings), in particular, the qualitative reconstructions related to Table \ref{tab:res1} where the training is performed using datasets obtained with $K=512$ angles and $I=10^4$ photons while in testing we have reconstructed CT images from sinograms containing either $256,512,1024$ angles. It is possible to clearly highlight how the RAN2I manages to preserve a better quality than N2I when the number of angles decreases in the reconstruction especially for a sparse number of angles as in Fig. \ref{fig:results_diff_proj}(b) which shows the case with 256 angles.

\subsubsection{Test with different geometry}

To study the generalization of RAN2I at different CT data acquisition geometry, we trained the network with the dataset obtained from parallel-beam and tested the model on the testing dataset generated from 2D fan-beam geometry. The simulated fan-beam detector array consists of 768 detectors with a width of 2 mm each. The distance from center of rotation to detector is 500 mm and source to center of rotation is 1000 mm. The noisy sinograms were obtained with 1024 angles and $10^4$ photons. FBP was used to reconstruct the sub-sinograms after the split.

\begin{table*}[!h]
	\caption{Quantitative results with $I_0=10^4$ and different number of angles with fan beam CT geometry.}\label{tab:res1fan}
	\begin{center}
		\setlength{\tabcolsep}{5pt} 
		\renewcommand{\arraystretch}{1.1} 
		\begin{tabular}{c|cccc||cccc}
			\hline \hline
            & \multicolumn{4}{c||}{Training: $K=512$ angles, $I_0 = 10^4$  photons} & \multicolumn{4}{c}{Training: $K=1024$ angles, $I_0 = 10^4$  photons} \\ 	\hline \hline
			angles & 128 & 256 & 512 & 1024 & 128 & 256 & 512 & 1024 \\ \hline 
			& \multicolumn{4}{c||}{PSNR} & \multicolumn{4}{c}{PSNR} \\ \hline
			RAN2I & $\mathbf{27.68}\pm 0.66$ & $\mathbf{32.56}\pm 0.65$ & $\mathbf{35.54}\pm 0.83$ & $\mathbf{36.39}\pm 0.91$ & $\mathbf{26.93}\pm 0.57$ & $\mathbf{31.23}\pm 0.48$ & $\mathbf{35.28}\pm 0.63$ & $\mathbf{37.23}\pm 0.87$  \\
			N2I & $27.30\pm 0.62$ & $31.56\pm 0.62$ & $34.43\pm 0.76$ & $35.90\pm 0.88$ & $26.88\pm 0.59$ & $31.15\pm 0.51$ & $34.87\pm 0.61$ & $36.90\pm 0.81$  \\ 
            Sup & $30.02\pm 0.47$ & $34.12\pm 0.45$ & $35.54\pm 0.60$ & $36.53\pm 0.62$ & $27.90\pm 0.41$ & $31.87\pm 0.48$ & $35.00\pm 0.87$ & $36.94\pm 0.63$ \\ \hline
			& \multicolumn{4}{c||}{SSIM} & \multicolumn{4}{c}{SSIM} \\ \hline
			RAN2I & $\mathbf{0.59}\pm 0.03$ & $\mathbf{0.83}\pm 0.02$ & $\mathbf{0.91}\pm 0.01$ & $\mathbf{0.92}\pm 0.01$ & $\mathbf{0.58}\pm 0.02$ & $\mathbf{0.78}\pm 0.01$ & $\mathbf{0.90}\pm 0.01$ & $\mathbf{0.93}\pm 0.01$ \\
			N2I & $0.56\pm 0.02$ & $0.80\pm 0.02$ & $0.90\pm 0.01$ & $0.92\pm 0.01$ & $0.56\pm 0.02$ & $0.77\pm 0.02$ & $0.88\pm 0.01$ & $0.92\pm 0.01$  \\
            Sup & $0.74\pm 0.01$ & $0.87\pm 0.01$ & $0.89\pm 0.01$ & $0.91\pm 0.01$ & $0.63\pm 0.01$ & $0.79\pm 0.02$ & $0.87\pm 0.01$ & $0.91\pm 0.01$ \\
			\hline \hline 
		\end{tabular}
	\end{center}
\end{table*}

\begin{table*}[!h]
	\caption{Quantitative results with the same angles and different amount of photon counts with fan beam CT geometry.}\label{tab:res2fan}
	\begin{center}
		\setlength{\tabcolsep}{5pt} 
		\renewcommand{\arraystretch}{1.1} 
		\begin{tabular}{c|cccc||cccc}
			\hline \hline
            & \multicolumn{4}{c||}{Training: $K=512$ angles, $I_0 = 10^4$  photons} & \multicolumn{4}{c}{Training: $K=1024$ angles, $I_0 = 10^4$  photons} \\ \hline \hline
			$I_0$ & $10^3$ & $5\cdot 10^3$ & $10^4$ & $10^5$ & $10^3$ & $5\cdot 10^3$ & $10^4$ & $10^5$ \\ \hline  
			& \multicolumn{4}{c||}{PSNR} & \multicolumn{4}{c}{PSNR} \\ \hline
			RAN2I & $\mathbf{28.20}\pm 0.40$ & $\mathbf{34.40}\pm 0.64$ & $\mathbf{35.54}\pm 0.83$ & $\mathbf{36.57}\pm 0.95$ & $\mathbf{29.50}\pm 0.44$ & $\mathbf{35.79}\pm 0.62$ & $\mathbf{37.23}\pm 0.87$ & $\mathbf{38.70}\pm 1.01$ \\
			N2I & $27.25\pm 0.43$ & $33.41\pm 0.62$ & $34.43\pm 0.76$ & $35.45\pm 0.85$ & $29.53\pm 0.42$ & $35.58\pm 0.89$ & $36.90\pm 0.81$ & $38.34\pm 0.0.97$ \\
            Sup & $27.95\pm 0.49$ & $34.24\pm 0.45$ & $35.54\pm 0.60$ & $38.68\pm 0.96$ & $28.15\pm 0.50$ & $35.02\pm 0.41$ & $36.94\pm 0.63$ & $39.96\pm 1.03$ \\
			\hline 
			& \multicolumn{4}{c||}{SSIM} & \multicolumn{4}{c}{SSIM} \\ \hline
			RAN2I & $\mathbf{0.65}\pm 0.02$ & $\mathbf{0.87}\pm 0.01$ & $\mathbf{0.91}\pm 0.01$ & $\mathbf{0.94}\pm 0.01$ & $\mathbf{0.71}\pm 0.02$ & $\mathbf{0.90}\pm 0.01$ & $\mathbf{0.93}\pm 0.01$ & $\mathbf{0.96}\pm 0.01$ \\
			N2I & $0.58\pm 0.03$ & $0.86\pm 0.01$ & $0.90\pm 0.01$ & $0.94\pm 0.01$ & $0.68\pm 0.02$ & $0.89\pm 0.01$ & $0.92\pm 0.01$ & $0.95\pm 0.01$ \\ 
            Sup & $0.68\pm 0.02$ & $0.87\pm 0.01$ & $0.89\pm 0.01$ & $0.95\pm 0.01$ & $0.67\pm 0.02$ & $0.87\pm 0.01$ & $0.91\pm 0.01$ & $0.96\pm 0.01$ \\
            \hline \hline 
		\end{tabular}
	\end{center}
\end{table*}

\begin{figure*}[!h]
	\centering
	\small\addtolength{\tabcolsep}{-18pt}
	\renewcommand{\arraystretch}{0.1}
	
	\begin{tabular}{ccccc}
		\hspace{-.6cm}\small{(a) Ground Truth} 
		& 
		\specialcell[c]{\hspace{-.3cm}\small (b) FBP}
		& 
		\specialcell[c]{\hspace{-.6cm}\small (c) N2I} 
		& 
		\specialcell[c]{\hspace{-.6cm}\small (d) RAN2I} 
		& 
		\specialcell[c]{\hspace{-.7cm}\small (e) Supervised}   \\
		\vspace{.2cm}
		\begin{tikzpicture}

			\begin{scope}[spy using outlines={rectangle,red,magnification=2,width=1cm, height=1.3cm,connect spies}]   
				\node {\includegraphics[viewport=0 10 400 360, clip, width=0.21\textwidth]{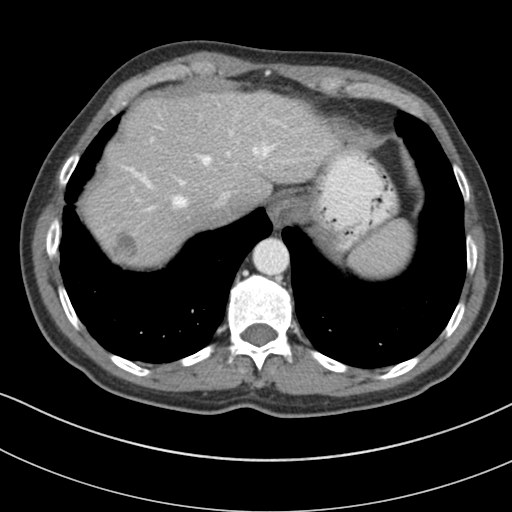}};
				\spy on (-1,.2) in node [left] at (-0.7,-1.8); 
				\draw [green, dashed] (.2,.5)--(.9,.5);
			\end{scope}
		\end{tikzpicture} \hspace{.25cm} 
		&
		\begin{tikzpicture}
			\begin{scope}[spy using outlines={rectangle,red,magnification=2,width=1cm, height=1.3cm,connect spies}]   
				\node {\includegraphics[viewport=0 10 400 360, clip, width=0.21\textwidth]{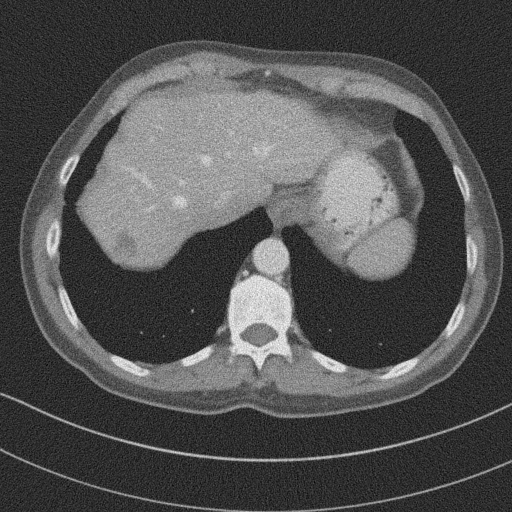}};
				\spy on (-1,.2) in node [left] at (-0.7,-1.8); 
			\end{scope}
		\end{tikzpicture} \hspace{.25cm} 
		&
		\begin{tikzpicture}
			\begin{scope}[spy using outlines={rectangle,red,magnification=2,width=1cm, height=1.3cm,connect spies}]   
				\node {\includegraphics[viewport=0 10 400 360, clip, width=0.21\textwidth]{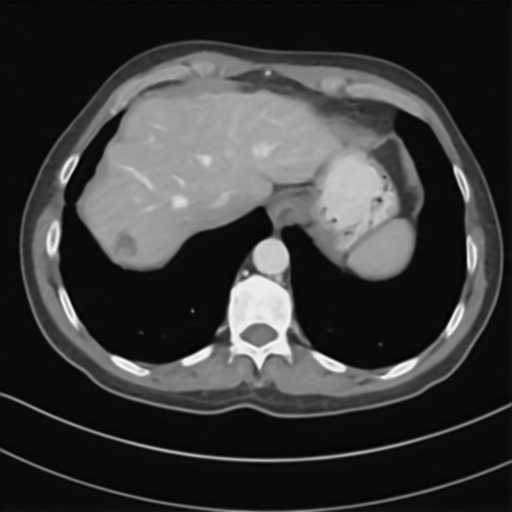}};
				\spy on (-1,.2) in node [left] at (-0.7,-1.8); 
			\end{scope}
		\end{tikzpicture} \hspace{.25cm} 
		&
		\begin{tikzpicture}
			\begin{scope}[spy using outlines={rectangle,red,magnification=2,width=1cm, height=1.3cm,connect spies}]   
				\node {\includegraphics[viewport=0 10 400 360, clip, width=0.21\textwidth]{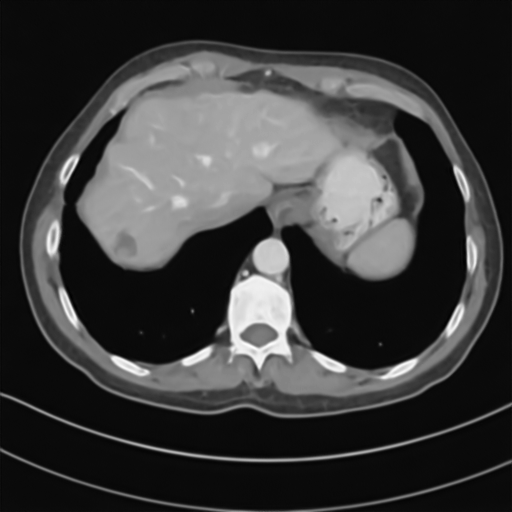}};
				\spy on (-1,.2) in node [left] at (-0.7,-1.8); 
			\end{scope}
		\end{tikzpicture}  \hspace{.25cm} 
		&       
		\begin{tikzpicture}
			\begin{scope}[spy using outlines={rectangle,red,magnification=2,width=1cm, height=1.3cm,connect spies}]   
				\node {\includegraphics[viewport=0 10 400 360, clip, width=0.21\textwidth]{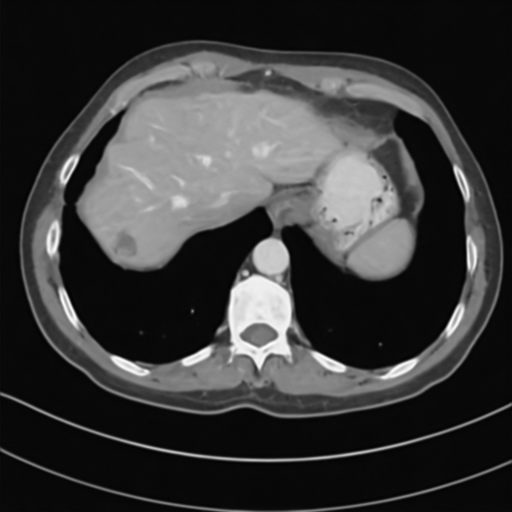}};
				\spy on (-1,.2) in node [left] at (-0.7,-1.8); 
			\end{scope}
		\end{tikzpicture}  
		\vspace{-.2cm} 
		\\ 
		\begin{tikzpicture}
			\node[text centered,text width = 0.15\textwidth, text depth = 3cm, anchor=north] {};
		\end{tikzpicture} 
		&
		\begin{tikzpicture}
			\begin{scope}[spy using outlines={rectangle,red,magnification=2,width=1cm, height=1.3cm,connect spies}]   
				\node {\includegraphics[viewport=0 10 400 360, clip, width=0.21\textwidth]{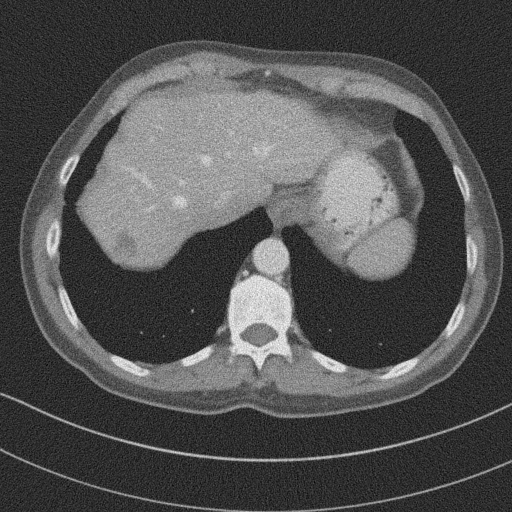}};
				\spy on (-1,.2) in node [left] at (-0.7,-1.8); 
			\end{scope}
		\end{tikzpicture} \hspace{.25cm} 
		&
		\begin{tikzpicture}
			\begin{scope}[spy using outlines={rectangle,red,magnification=2,width=1cm, height=1.3cm,connect spies}]   
				\node {\includegraphics[viewport=0 10 400 360, clip, width=0.21\textwidth]{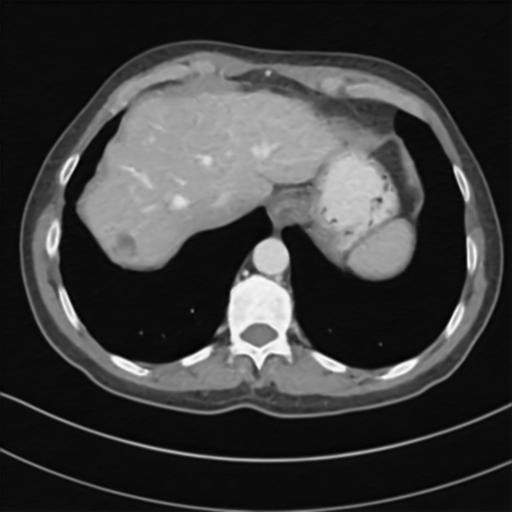}};
				\spy on (-1,.2) in node [left] at (-0.7,-1.8); 
			\end{scope}
		\end{tikzpicture}  \hspace{.25cm} 
		&       
		\begin{tikzpicture}
			\begin{scope}[spy using outlines={rectangle,red,magnification=2,width=1cm, height=1.3cm,connect spies}]   
				\node {\includegraphics[viewport=0 10 400 360, clip, width=0.21\textwidth]{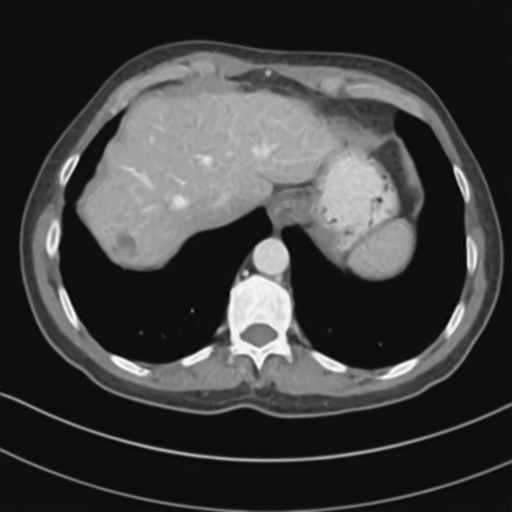}};
				\spy on (-1,.2) in node [left] at (-0.7,-1.8); 
			\end{scope}
		\end{tikzpicture} \hspace{.25cm} 
		&                             
		\begin{tikzpicture}
			\begin{scope}[spy using outlines={rectangle,red,magnification=2,width=1cm, height=1.3cm,connect spies}]
				\node {\includegraphics[viewport=0 10 400 360, clip, width=0.21\textwidth]{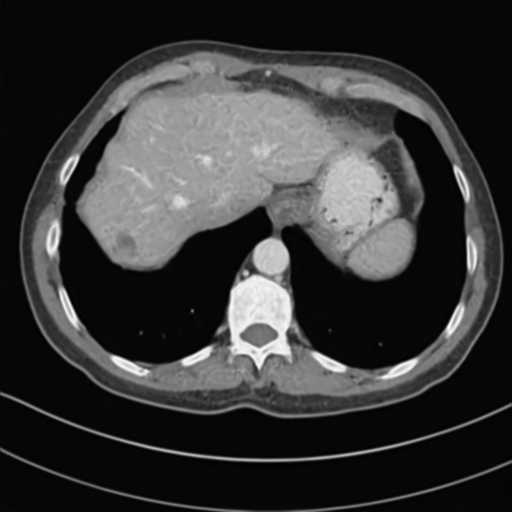}};
				\spy on (-1,.2) in node [left] at (-0.7,-1.8); 
			\end{scope}
		\end{tikzpicture}   
		
	\end{tabular}

	\caption{Qualitative reconstruction with $K= 1024$ angles and with different CT acquisition geometries, (top row) parallel beam and (bottom row) fan beam: a) represents the ground truth image, b) FBP, c) N2I, d) RAN2I proposed algorithm and e) reconstruction with supervised architecture. The HU display window is $[-450, 550]$ to highlight the lesion area.} \label{fig:geom_results}
\end{figure*}
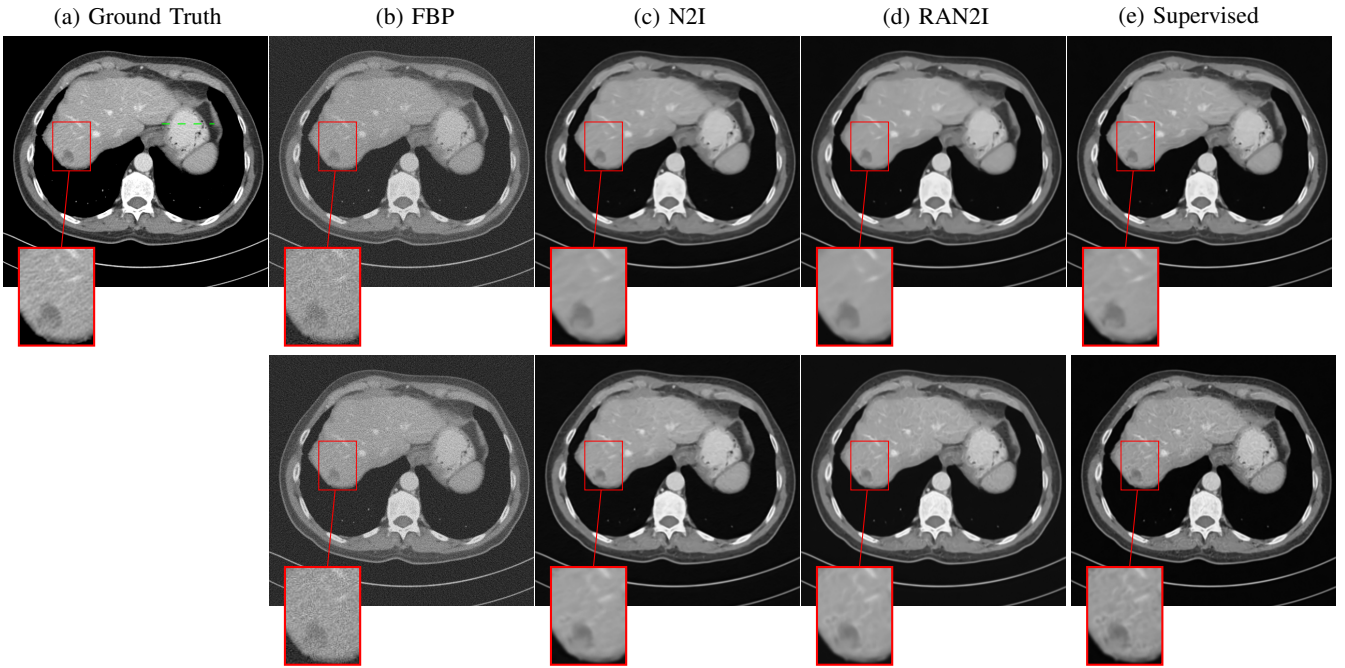

Table \ref{tab:res1fan} shows the mean PSNR and SSIM with the standard deviation of RAN2I, N2I, and supervised method, tested with $I_0=10^4$ and different number of angles with fan-beam geometry when the network is trained with 512 or 1024 angles and $I_0=10^4$ with parallel-beam acquisition (testing with mismatched geometry). RAN2I outperforms N2I at all sampling angles when tested with fan-beam geometry. 
Similar results can be observed in Table \ref{tab:res2fan} which shows the quantitative results with the same angles and different photon counts. It is noticeable that the result values of supervised method under mismatched geometry decrease dramatically. 

\begin{figure*}[!h]
	
	\begin{minipage}{.48\textwidth}
		\centering
		\subfloat[Parallel beam CT acquisition geometry]{
			\begin{tikzpicture}[scale=1, spy using outlines={rectangle,purple,magnification=1.5,size=15mm,connect spies}] 
				\begin{axis}[
					xlabel={Distance (in pixels)},
					ylabel={Attenuation (mm$^{-1}$)},
					xtick={100, 200, 300, 400, 500},
					grid = major,
					xmin=300, xmax=400,
					ymin=0.17, ymax=.35,
					legend columns=2,
					legend cell align=left,
					legend entries={Ground truth, RAN2I, N2I, Supervised},
					legend style={at={(0.4, 0.2)},anchor=north},
					width=\textwidth,
					height=\axisdefaultheight
					]				
					\addplot [color=blue] table [x=x_axis, y=groundtruth, col sep=comma]{Figure8a.txt};	
					\addplot [color=red] table [x=x_axis, y=EN2I, col sep=comma]{Figure8a.txt};	
					\addplot [color=green] table [x=x_axis, y=N2I, col sep=comma]{Figure8a.txt};	
					\addplot [color=black] table [x=x_axis, y=supervised, col sep=comma]{Figure8a.txt};	
				\end{axis}
				\spy on (2.5,3.1) in node [left] at (5.5,4.5); 
		\end{tikzpicture}}
	\end{minipage}
	\begin{minipage}{.48\textwidth}
		\centering
		\subfloat[Fan beam CT acquisition geometry]{
			\begin{tikzpicture}[scale=1, spy using outlines={rectangle,purple,magnification=1.5,size=15mm,connect spies}] 
				\begin{axis}[
					xlabel={Distance (in pixels)},
					ylabel={Attenuation (mm$^{-1}$)},
					xtick={100, 200, 300, 400, 500},
					xticklabels={100, 200, 300, 400, 500},
					grid = major,
					xmin=300, xmax=400,
					ymin=0.17, ymax=.35,
					legend columns=2,
					legend cell align=left,
					legend entries={Ground truth, RAN2I, N2I, Supervised},
					legend style={at={(0.4, 0.2)},anchor=north},
					width=\textwidth,
					height=\axisdefaultheight
					]				
					\addplot [color=blue] table [x=x_axis, y=groundtruth, col sep=comma]{Figure8b.txt};	
					\addplot [color=red] table [x=x_axis, y=EN2I, col sep=comma]{Figure8b.txt};	
					\addplot [color=green] table [x=x_axis, y=N2I, col sep=comma]{Figure8b.txt};	
					\addplot [color=black] table [x=x_axis, y=supervised, col sep=comma]{Figure8b.txt};	
				\end{axis}
				\spy on (2.5,3.1) in node [left] at (5.5,4.5);
		\end{tikzpicture}}
		
	\end{minipage}
	
	\caption{Comparison of the estimated attenuation coefficients along the green line in the ground truth image in Fig. \ref{fig:geom_results} using a) parallel and b) fan beam CT geometries.} \label{fig:int_plot_geom}
	
\end{figure*}
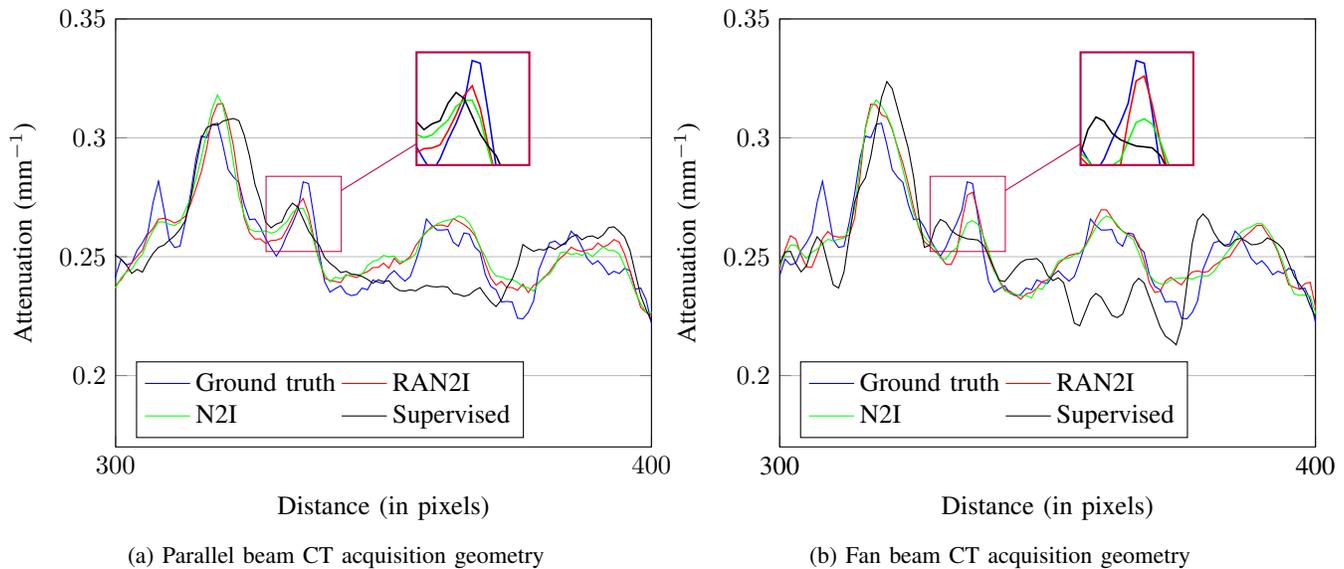

Fig. \ref{fig:geom_results} compares the output images obtained with parallel-beam and fan-beam geometry. We can observe that the results of fan-beam geometry are worse than that of parallel-beam geometry in the testing, as the training datasets are obtained with only parallel-beam geometry. However, the supervised method shows worse generalization ability than both N2I and RAN2I, as the result for fan-beam geometry is more noisy. Fig. \ref{fig:int_plot_geom} compares the attenuation along the green line of the images in Fig. \ref{fig:geom_results} and the lines of RAN2I are closer to the ground truth than N2I in both geometries. In addition, the result of supervised method deviates more from ground truth when testing with fan-beam geometry.

\begin{figure*}[!b]
    \begin{minipage}{.32\textwidth}
    \subfloat[Full-dose]{
    \centering
    \includegraphics[width=.88\textwidth]{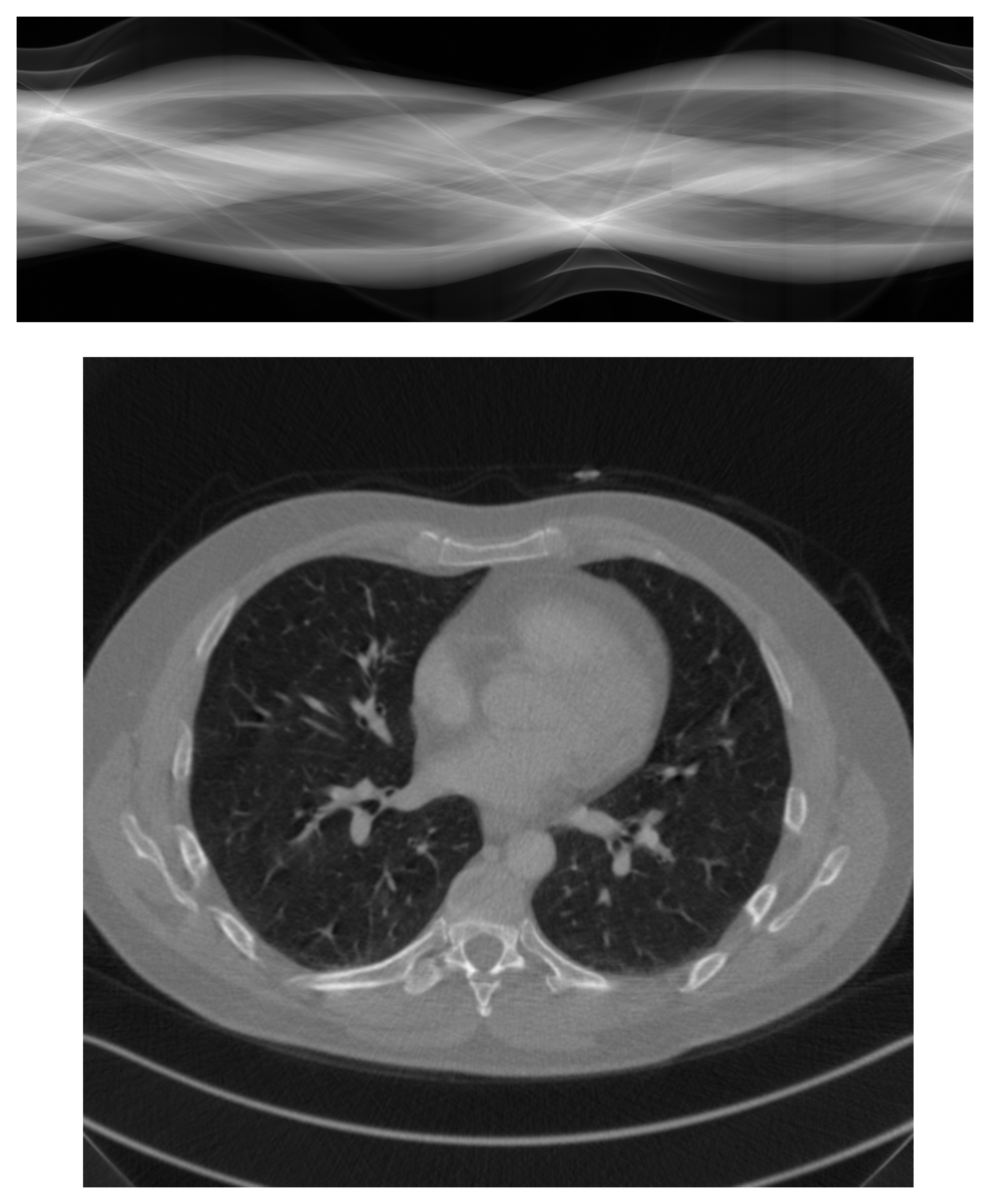}}
    \end{minipage}
    \begin{minipage}{.32\textwidth}
    \subfloat[Quarter-dose]{
    \includegraphics[width=.88\textwidth]{Figure9b.png} }
    \end{minipage}
    \begin{minipage}{.32\textwidth}
    \subfloat[Split quarter-dose]{
    \centering
    \includegraphics[width=.88\textwidth]{Figure9c.png} }
    \end{minipage}
    \caption{Examples of split re-binned projection and reconstruction. a) Full-dose projection and FBP reconstruction. b) Quarter-dose projection and FBP reconstruction. c) Split quarter-dose projection and FBP reconstruction.}
    \label{fig:split_example_raw}
\end{figure*}

\section{Experimental Results}\label{sec: experimental results}

\subsection{Data processing}

In addition to numerical simulations, we also conducted real-data experiments, applying the proposed method directly to projection data from the 2016 NIH-AAPM-Mayo Clinic Low Dose CT Grand Challenge \cite{moen2021low} which also provides full-dose and quarter-dose helical projection data. We used the Helix2Fan \cite{wagner2022dual} algorithm for re-binning helical projections into 2D fan-beam geometry. Then we used ASTRA Toolbox \cite{van2015astra} to reconstruct re-binned projections based on the geometry settings of real acquisition, in which the detector counts are 736 and each detector is 1.84 mm wide; sampling angles are 2304; distance from origin to source is 850 mm and distance from origin to detector is 700.86 mm. 

To obtain the training dataset for RAN2I, the noisy quarter-dose projections were split angularly into two subsets, and the geometry was also split to reconstruct sub-sinograms using FBP. Finally, 308 slices of noisy reconstructed images from 5 patients were used as training datasets, and 22 slices of noisy reconstructed images from another 1 patient were used as testing datasets for the RAN2I experiment. Fig. \ref{fig:split_example_raw} shows an example of split re-binned projection data and the associated FBP reconstruction.

\begin{figure*}[!h]
	\centering
	\small\addtolength{\tabcolsep}{-18pt}
	\renewcommand{\arraystretch}{0.1}
	
	\begin{tabular}{ccccc}
		\hspace{-.6cm}\small{(a) Full dose} 
		& 
		\specialcell[c]{\hspace{-.6cm}\small (b) Quarter dose}
		& 
		\specialcell[c]{\hspace{-.6cm}\small (c) RAN2I} 
        & 
		\specialcell[c]{\hspace{-.7cm}\small (d) Supervised (308 slices)}
		& 
		\specialcell[c]{\hspace{-.7cm}\small (e) Supervised}   \\
		\vspace{.2cm}
		\begin{tikzpicture}
			\begin{scope}[spy using outlines={rectangle,red,magnification=3,width=2cm, height=1.5cm,connect spies}]   
				\node {\includegraphics[viewport=0 10 400 360, clip, width=0.21\textwidth]{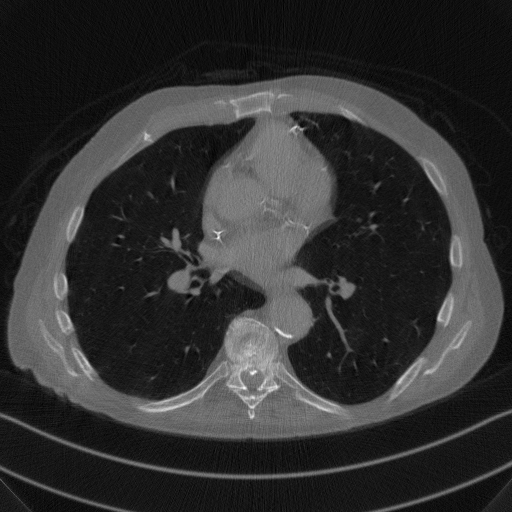}};
				\spy on (-.15,-.9) in node [left] at (-0.0,-2); 
				\draw [green,dashed] (-1.77,0)--(1.5,0);
			\end{scope}
		\end{tikzpicture} \hspace{.25cm}
		&
		\begin{tikzpicture}
			\begin{scope}[spy using outlines={rectangle,red,magnification=3,width=2cm, height=1.5cm,connect spies}]   
				\node {\includegraphics[viewport=0 10 400 360, clip, width=0.21\textwidth]{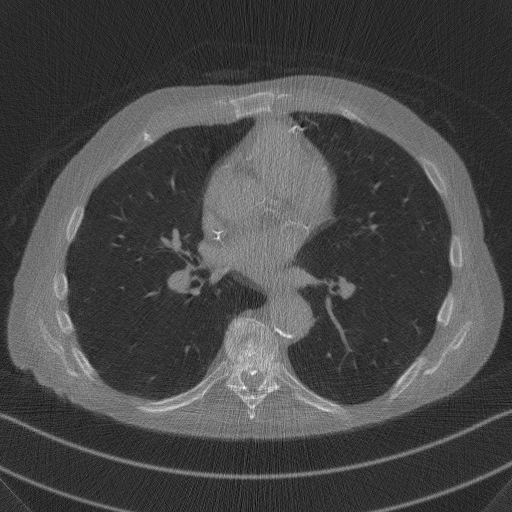}};
				\spy on (-.15,-.9) in node [left] at (-0.0,-2); 
			\end{scope}
		\end{tikzpicture} \hspace{.25cm}
		&
		\begin{tikzpicture}
			\begin{scope}[spy using outlines={rectangle,red,magnification=3,width=2cm, height=1.5cm,connect spies}]   
				\node {\includegraphics[viewport=0 10 400 360, clip, width=0.21\textwidth]{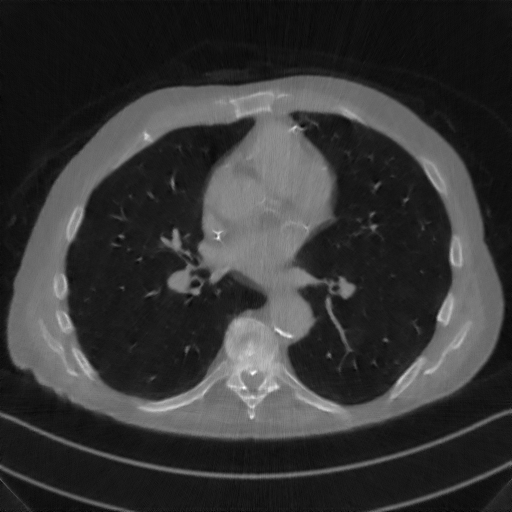}};
				\spy on (-.15,-.9) in node [left] at (-0.0,-2); 
			\end{scope}
		\end{tikzpicture} \hspace{.25cm}
        &       
        \begin{tikzpicture}
			\begin{scope}[spy using outlines={rectangle,red,magnification=3,width=2cm, height=1.5cm,connect spies}]   
				\node {\includegraphics[viewport=0 10 400 360, clip, width=0.21\textwidth]{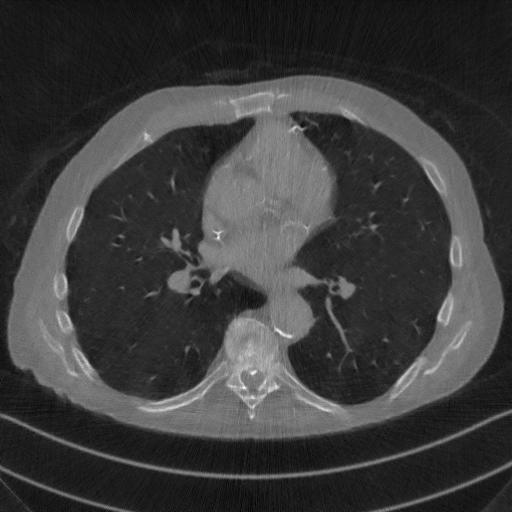}};
				\spy on (-.15,-.9) in node [left] at (-0.0,-2); 
			\end{scope}
		\end{tikzpicture} \hspace{.25cm}
		&       
        \begin{tikzpicture}
			\begin{scope}[spy using outlines={rectangle,red,magnification=3,width=2cm, height=1.5cm,connect spies}]   
				\node {\includegraphics[viewport=0 10 400 360, clip, width=0.21\textwidth]{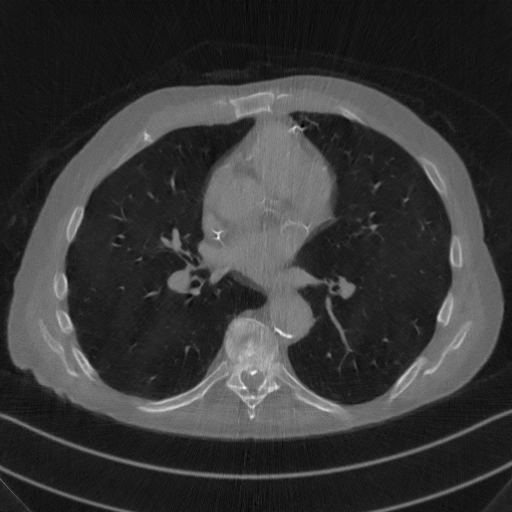}};
				\spy on (-.15,-.9) in node [left] at (-0.0,-2); 
			\end{scope}
		\end{tikzpicture} \hspace{.25cm}
		\\   
		
	\end{tabular}
	\caption{Experimental reconstruction: a) Full dose FBP, b) Quarter dose FBP, c) RAN2I, d) Supervised with 308 training slices, e) Supervised with 649 training slices.} \label{fig:experimental results}
\end{figure*}

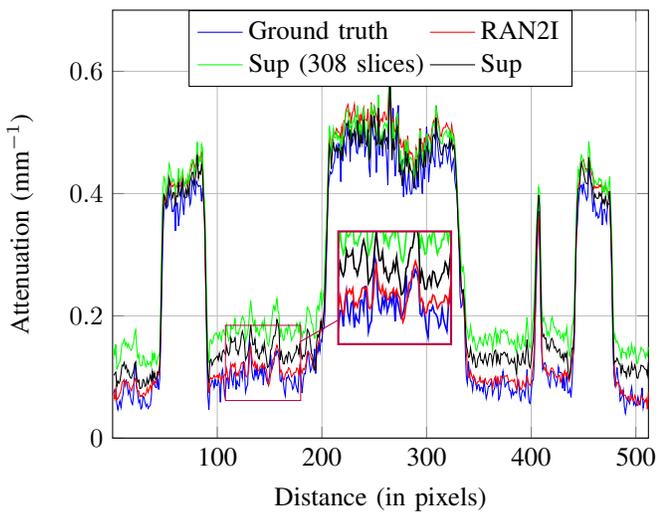
\begin{figure}[!h]
	
	\begin{minipage}{.48\textwidth}
		\centering
		\subfloat[Parallel beam CT acquisition geometry]{
			\begin{tikzpicture}[scale=1, spy using outlines={rectangle,purple,magnification=1.5,size=15mm,connect spies}] 
				\begin{axis}[
					xlabel={Distance (in pixels)},
					ylabel={Attenuation (mm$^{-1}$)},
					xtick={100, 200, 300, 400, 500},
					xticklabels={100, 200, 300, 400, 500},
					grid = major,
					xmin=0, xmax=512,
					ymin=0, ymax=.7,
					legend columns=2,
					legend cell align=left,
					legend entries={Ground truth, RAN2I, Sup (308 slices), Sup},
					legend style={at={(0.5, 1)},anchor=north},
					width=\textwidth,
					height=\axisdefaultheight
					]				
					\addplot [color=blue] table [x=x_axis, y=groundtruth, col sep=comma]{Figure11.txt};	
					\addplot [color=red] table [x=x_axis, y=RAN2I, col sep=comma]{Figure11.txt};	
					\addplot [color=green] table [x=x_axis, y=sup_half, col sep=comma]{Figure11.txt};	
					\addplot [color=black] table [x=x_axis, y=supervised, col sep=comma]{Figure11.txt};	
				\end{axis}
				\spy on (2,1) in node [left] at (4.5,2); 
		\end{tikzpicture}}
	\end{minipage}
	
	\caption{Comparison of the estimated attenuation coefficients along the green line in the ground truth image in Fig. \ref{fig:experimental results}.
 } \label{fig:int_plot_raw}
	
\end{figure}

\subsection{Comparison with different methods}

RAN2I was compared with the supervised method in the experiment. The quarter-dose projections and full-dose projections were directly reconstructed using FBP and the reconstructed quarter-dose images and full-dose images were used as input and target respectively in the training. 649 slices from 8 patients were used as training datasets and 22 slices from another 1 patient were used as testing datasets. The training settings and the network architecture were the same as described in section \ref{sec:Setups}, except that the training epoch was set to 200. In addition, we also ran the supervised training with the same amount of slices as RAN2I which were 308, to show how the number of training data affect the final results.

\subsection{Results}

Fig. \ref{fig:experimental results} shows the results of RAN2I and supervised experiments. It can be noticed that RAN2I removes the artifacts effectively and has less noise. Though supervised results are closer to the full-dose image, the artifacts in the zoomed area are more than that in the RAN2I result, because the supervised method depends heavily on the target in the training, which is the full-dose image. However, in this case, the full-dose image is not the accurate ground truth. Also, the overall contrast of the supervised method is worse than that of the RAN2I, especially when the training datasets for supervised are only 308 slices which is the same as RAN2I. This indicates that RAN2I can achieve decent results with less training data and time compared to the supervised method.

Fig. \ref{fig:int_plot_raw} shows the intensity plot along the green line in Fig. \ref{fig:experimental results}. The line profile and the whole images in Fig. \ref{fig:experimental results} show that RAN2I has better overall contrast. The zoomed part of the line profile shows that RAN2I has the more accurate result. However, the high-intensity area shows that RAN2I has higher intensity due to the smoothness of artifacts while the supervised method has similar intensity with the full-dose image but preserves more noise and artifacts.

\section{Discussion}\label{ch5_discuss} 

\subsection{Parameters settings}

We performed all the numerical simulations using a $S=2$ split of the sinogram as the final output of the N2I method is based on the mean values of pixels and increasing split will result in more blurry results and evidence in \cite{hendriksen2020noise2inverse} shows that adding more splits does not significantly improve the results and even degrades the performance. It is worth mentioning that the type of transformation depends on the specific design of the imaging systems; for instance, rotation equivariance is embedded within how the CT scanning is performed.

In our work, we focus on LDCT reconstruction using two modalities to reduce X-ray dose, either by reducing the tube current or by reducing sampling views per scan. We will discuss the improvement and limitations of the proposed method RAN2I.

\subsection{Poisson noise}

Reducing the X-ray dose in CT, can be achieved by decreasing the X-ray tube current, i.e. the number of emitted photons. Poisson noise is one of the major photon noise in LDCT imaging due to insufficient photons. In Poisson noise experiments, RAN2I shows better performance than both N2I and score-based diffusion model, and is comparable to supervised method. Regarding N2I, the problem of is that with low photon counts the assumption of noise independence on the FBP images from sub-sampled sinogram is degraded and in fact, the results show that our proposed rotational data augmentation procedure RAN2I can improve substantially. Score-based model is designed to solve sparse-view inverse problem and cannot generalize well with the presence of Poisson noise.

\subsection{Sparse-view CT}

The other way to reduce X-ray dose is to reduce the sampling angles per scan. However, in the sparse view CT case the split of the sinogram results in solving a highly under-determined system. This limitation is inherent both for the N2I and RAN2I framework. In the sparse-view CT simulations, score-based diffusion model shows state-of-the-art performance in removing artifacts and outperforms supervised method, thanks to its powerful generative ability. N2I, RAN2I and supervised methods produce more blurry results but supervised method has better results in removing artifacts. In the sparse view results, we show that RAN2I is able to outperform N2I but not able to achieve the performance of the diffusion model. Furthermore, results have proven that RAN2I has better generalization than N2I in a wide range of sampling angles.

\subsection{Comparison with score-based generative model}

The results section \ref{sec: numerical results} presented a comprehensive analysis of RAN2I compared with state-of-the-art unsupervised method score-based generative model both in the sparse-view CT and low photon counts acquisition scenarios. 

In the sparse view CT simulation, RAN2I cannot achieve the accuracy of the score-based diffusion model \cite{song2021solving} which has a remarkably high baseline. The score-based generative model can remove the artifacts in sparse-view CT image due to its powerful generative ability. The framework of \cite{song2021solving} is constructed on the measurement sub-sampling acquisition model $\*A = \*S\*P$ where $\*P$ is the full forward operator and $\*S$ is a measurement sub-sampling operator. This scenario corresponds exactly to the sparse-view CT model. The diffusion model generates CT images consistent with the measurements $\*y$, through the up-sampling operator $\*S^{-1}$ and back-projection $\*P^{-1}$ but does not consider the constraint on the CT noise model. 

However, in the low photons acquisition which can be modelled with a Poisson noise statistics in the measurement domain, the score-based diffusion model noticeably under-perform compared to RAN2I. The conditional constraint of the score-based model does not account for the noise in the measurements $\*y$, therefore the diffusion model is not consistent with the CT noise model. While in the sparse view case this can be neglected if we consider normal dose, in the case of low-dose Poisson statistics, where the noise component highly deteriorate the quality of the image, it is necessary to take into account the CT noise statistics into the diffusion model. This is the reason of the noticeable degradation of the results of \cite{song2021solving} in case of low-dose CT for our comparison. 

On the contrary, RAN2I can remove Poisson noise effectively but cannot gain much improvement in sparse-view CT. The underlying strategy of RAN2I is based on Noise2Noise \cite{lehtinen2018noise2noise} method where the input and target in end-to-end training are two same images with different noise distribution. This process can be approximated to an end-to-end supervised training strategy as described in section \ref{subsec:N2I}. The objective of RAN2I is to output the conditional expectation given a noisy image, which is more like removing the additive noise in the image instead of generating a similar image like score-based generative model does. However, the baseline of RAN2I for sparse-view CT cannot reach the height of score-based generative model as the underlying strategy of RAN2I is still based on image domain. 

In terms of the clinical impact, RAN2I has more advantages than score-based model. The overall workflow of RAN2I is simple and the size of network is much smaller than that in score-based generative model. Therefore, the computational cost of RAN2I is much lower compared to score-based model. In addition, as an image-based method, the computational time of RAN2I is similar to FBP which can achieve real-time imaging. While the inference time of score-based model is slow and it takes couple minutes to generate one CT image.

\subsection{Generalization}

RAN2I shows better generalization than N2I for different CT acquisition settings and geometries. When the sampling angles are fewer, RAN2I yields better results than N2I. Because in the N2I framework, training datasets are obtained from split sinograms, which means the quality of the split data in training is worse than the original acquisition as shown in Fig. \ref{fig:split_example} and Fig. \ref{fig:split_example_raw}. Therefore, RAN2I is proposed to reduce this negative effect by exploiting additional information and the results prove its efficiency.

Though RAN2I loses to the supervised method in numerical simulations in some cases, in experimental results, RAN2I shows its advantages over the supervised method. The proposed method is tested on real projection data; only quarter-dose projections are required. Only 308 reconstructed slices are used in the training for RAN2I, while with the same amount of training slices, the supervised method produces worse results. The supervised method with 649 training slices produces results that are close to the full-dose target image but requires more full-dose datasets and more training epochs. In conclusion, the supervised method depends heavily on the target image and RAN2I is a better method when the accurate ground truth is unavailable, e.g. in our experiment, the rebinned reconstructed full-dose image deviates from the accurate result of vendor scanners.

\subsection{Limitations}

Our method is inspired by EI \cite{chen2021equivariant} to use equivariance property for data augmentation. However, our method does not exploit the exact equivariance defined in EI, as the training strategy we employ is based on the N2I framework where we enforce equivariance such that the output of the network $f_{\*\varphi}$ after applying the group transformation $T_g$ is the same as the target after imposing the same transformation. Compared to using equivariance within contrastive learning \cite{dangovski2021equivariant}, \cite{liu2021self}, our RAN2I method innovates in the way that equivariance is enforced in the image domain through the training loss inspired from N2I. Respect to \cite{laine2019high} where the loss function is designed from a probabilistic derivation, in RAN2I the loss in \eqref{eq:opt_eqn2i} is defined through a deterministic optimization problem to promote the rotational equivariance. The image-based augmentation strategy in RAN2I is similar to that in \cite{desai2021vortex} which proposes both physics-driven augmentation and image-based augmentation for accelerated MRI reconstruction but in our work, we do not implement physics-driven augmentation. Our initial thought is to incorporate physics-driven equivariance into the N2I framework to further improve the result of sparse views acquisition, combining the strict EI method with the N2I framework. However, this strategy has high computational cost and the re-sampling process in EI heavily degrades the overall performance as the training datasets will be corrupted by more artifacts and noise.

There are also several possible improvements in RAN2I. First of all, we only employ the equivariance term in the image domain, which does not improve the results in very sparse views. To address this problem, our plan for the next stage is to consider equivariance in the projection domain, referring to \cite{choi2023self}. Secondly, in the very sparse view case, blurriness is a major problem in the N2I framework, as the output is the mean of split images, plus the MSE loss function contributes to blurriness. Modification can be done by changing the basic framework or designing a specific loss function, which is left for future work.

\section{Conclusion}

We proposed a new self-supervised image denoising method RAN2I for LDCT by enforcing equivariant rotation using a novel training loss function for the  N2I framework. EN2I improves both quantitative the qualitative results compared with the existing N2I method. In addition, the reconstructed image of RAN2I is close to supervised method and ground truth is not required for the training, which is crucial for LDCT reconstruction in clinical applications.

\section*{Acknowledgement}

All authors declare that they have no known conflicts of interest in terms of competing financial interests or personal relationships that could have an influence or are relevant to the work reported in this paper. 

\appendix

\subsection{Prediction error of the RAN2I algorithm} \label{app_conv}

\begin{proof}[Proposition \ref{prop:pred_error}]
By defining 
\begin{align}
   \*y &= \*y^* + \bm\epsilon & \*y^* &= \*A\*x \nonumber \\
   \*x^*_J &= \*R_J\*y^*_J & \*x^*_{J^C} &= \*R_{J^C}\*y^*_{J^C}
\end{align}
following arguments from \cite{hendriksen2020noise2inverse}, we can expand the squared norm of the error between the equivariant output and clean reference
\begin{align}
    \| & T_g f_{\*\varphi}(\hat{\*x}_{J^C}) - T_g (\hat{\*x}_{J}) \|^2\nonumber \\
    & = \| T_g f_{\*\varphi}(\hat{\*x}_{J^C}) - T_g (\*x^*_{J}) + T_g (\*x^*_{J}) - T_g (\hat{\*x}_{J}) \|^2 \nonumber \\
    & = \| T_g f_{\*\varphi}(\hat{\*x}_{J^C}) - T_g (\*x^*_{J}) \|^2 + \| T_g (\*x^*_{J}) - T_g (\hat{\*x}_{J}) \|^2 \nonumber \\
    & \quad + 2\langle T_g f_{\*\varphi}(\hat{\*x}_{J^C}) - T_g (\*x^*_{J}), T_g (\*x^*_{J}) - T_g (\hat{\*x}_{J}) \rangle
\end{align}
Given $\*y^* = \*A\*x$ and $J\in\-J$, using (\ref{eq:expectation}) we obtain
\begin{eqnarray}
    \@E_{x, \epsilon}\left[ T_g (\hat{\*x}_{J}) |\, \*x, J \right] &=& \@E_{x, \epsilon}\left[ T_g (\*R_J \*y_J) | \, \*x, J \right] \nonumber \\ 
    & = & T_g \*R_J \@E_{x, \epsilon}\left[ \*y^*_J + \bm\epsilon_J |\, \*x \right] \nonumber \\
    &=& T_g \*R_J\*y^*_J = T_g \*x^*_J
\end{eqnarray}
The noisy random variables $\hat{\*x}_{J^C}$ and $\hat{\*x}_J$ are independent conditioned on $\*x$ and $J$, since the domains of $\*R_J$ and $\*R_{J^C}$ do not overlap, and the noise $\bm\epsilon$ is element-wise statistically independent. This independence condition allows us to interchange the order of the expectation and inner product which yields using (\ref{eq:expectation})
\begin{align}
	\@E &\left[ \langle T_g f_{\*\varphi}(\hat{\*x}_{J^C}) - T_g (\*x^*_J), T_g (\*x^*_J) - T_g (\hat{\*x}_J) |\, \*x, J \rangle \right] \nonumber \\
	& = \langle \@E \left[T_g f_{\*\varphi}(\hat{\*x}_{J^C}) - T_g (\*x^*_J)|\, \*x, J \right], \nonumber \\
 & \quad \quad \quad \;\, \@E \left[T_g (\*x^*_J) - T_g (\hat{\*x}_J) |\, \*x, J \right] \rangle \nonumber \\ 
	& = \langle \@E \left[T_g f_{\*\varphi}(\hat{\*x}_{J^C}) - T_g (\*x^*_J) |\, \*x, J \right], 0\rangle = 0
\end{align}
Using the tower property of expectation, we obtain
\begin{align}
& \@E_{x,\epsilon} \| T_g f_{\*\varphi}(\hat{\*x}_{J^C}) - T_g (\hat{\*x}_{J}) \|^2  \\
& = \@E \left[\@E \| T_g f_{\*\varphi}(\hat{\*x}_{J^C}) - T_g (\hat{\*x}_{J}) \|^2  |\, \*x, J\right] \nonumber \\
& = \@E \left[\@E \left[\| T_g f_{\*\varphi}(\hat{\*x}_{J^C}) - T_g ({\*x}^*_J) \|^2 + \| T_g ({\*x}^*_J) - T_g (\hat{\*x}_{J}) \|^2 \right] 
\right] \nonumber \\
& = \@E_{x,\epsilon} \| T_g f_{\*\varphi}(\hat{\*x}_{J^C}) - T_g ({\*x}^*_J) \|^2 + \@E_{x,\epsilon} \| T_g ({\*x}^*_J) - T_g (\hat{\*x}_{J}) \|^2 \nonumber
\end{align}
A similar derivation can be used to obtain the relation 
\begin{align}
\@E_{x,\epsilon} & \| f_{\*\varphi}(\hat{\*x}_{J^C}) - (\hat{\*x}_{J}) \|^2  \\
	& = \@E_{x,\epsilon} \| f_{\*\varphi}(\hat{\*x}_{J^C}) - ({\*x}^*_J) \|^2 + \@E_{x,\epsilon} \| ({\*x}^*_J) - (\hat{\*x}_{J}) \|^2 \nonumber
\end{align}
\end{proof}

\subsection{Analysis of hyper-parameters}

To study the effect of hyper-parameters, we trained the network with different group transformation settings, including selecting random rotations or fixed rotations at every training batch for all the datasets, and number of rotations $r$. 

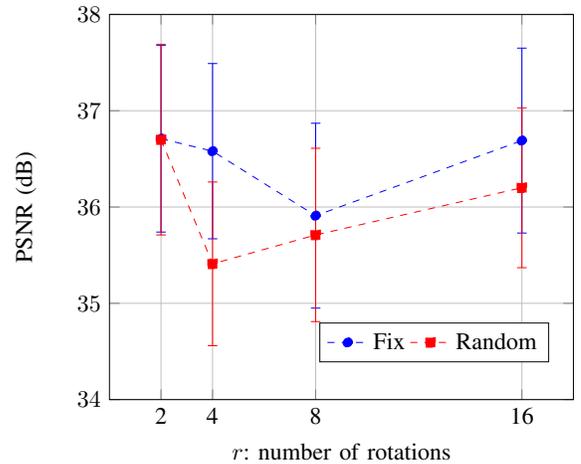
\begin{figure}[!h]
	\centering
		\begin{tikzpicture}[scale=.9] 
			\begin{axis}[
				xlabel={$r$: number of rotations},
				ylabel={PSNR (dB)},
				grid = major,
				xtick={2, 4, 8, 16},
				xticklabels={2, 4, 8, 16},
				xmin=0, xmax=18,
				ymin=34, ymax=38,
				legend columns=2,
				legend cell align=left,
				legend entries={Fix, Random},
				legend style={at={(0.7,0.2)},anchor=north},
				]				
				\addplot[color=blue, mark=*, style=dashed, error bars/.cd, y explicit, y dir=both, error bar style={color=blue,solid}] table[x=x_axis, y=fix, y error plus expr=\thisrow{fix_std},
				y error minus expr=\thisrow{fix_std}] {Figure12.txt};
				\addplot[color=red, mark=square*, style=dashed, error bars/.cd, y explicit, y dir=both, error bar style={color=red,solid}] table[x=x_axis, y=Random, y error plus expr=\thisrow{Random_std},
y error minus expr=\thisrow{Random_std}] {Figure12.txt};		
			\end{axis}
		\end{tikzpicture}
	\caption{Analysis of the reconstruction accuracy of the RAN2I algorithm with respect to different number of rotations considered in the rotation loss and whether the rotation is fixed or random during training.} \label{fig:n_rotations}
	
\end{figure}

The network was trained with $r=2, 4, 8, 16$ random or fixed rotations respectively to compare the denoising performance of different configurations. All the trainings above were performed on $K=512$ angles datasets.
Fig. \ref{fig:n_rotations} compares the denoising performance of random and fixed rotations as well as different numbers of rotations in the testing.

The line graph shows both random and fixed rotation has the highest PSNR when the number of rotations $r=2$ and the fixed rotation has better performance with increasing number of rotations. 

\bibliographystyle{IEEEtran}
\bibliography{biblio}

\begin{thebibliography}{10}
\providecommand{\url}[1]{#1}
\csname url@samestyle\endcsname
\providecommand{\newblock}{\relax}
\providecommand{\bibinfo}[2]{#2}
\providecommand{\BIBentrySTDinterwordspacing}{\spaceskip=0pt\relax}
\providecommand{\BIBentryALTinterwordstretchfactor}{4}
\providecommand{\BIBentryALTinterwordspacing}{\spaceskip=\fontdimen2\font plus
\BIBentryALTinterwordstretchfactor\fontdimen3\font minus
  \fontdimen4\font\relax}
\providecommand{\BIBforeignlanguage}[2]{{%
\expandafter\ifx\csname l@#1\endcsname\relax
\typeout{** WARNING: IEEEtran.bst: No hyphenation pattern has been}%
\typeout{** loaded for the language `#1'. Using the pattern for}%
\typeout{** the default language instead.}%
\else
\language=\csname l@#1\endcsname
\fi
#2}}
\providecommand{\BIBdecl}{\relax}
\BIBdecl

\bibitem{venkatakrishnan2013plug}
S.~V. Venkatakrishnan, C.~A. Bouman, and B.~Wohlberg, ``Plug-and-play priors
  for model based reconstruction,'' in \emph{2013 IEEE Global Conference on
  Signal and Information Processing}.\hskip 1em plus 0.5em minus 0.4em\relax
  IEEE, 2013, pp. 945--948.

\bibitem{hu2011lowrank}
Y.~Hu, S.~G. Lingala, and M.~Jacob, ``A fast majorize–minimize algorithm for
  the recovery of sparse and low-rank matrices,'' \emph{IEEE Transactions on
  Image Processing}, vol.~21, no.~2, pp. 742--753, 2011.

\bibitem{niu2014}
S.~Niu, Y.~Gao, Z.~Bian, J.~Huang, W.~Chen, G.~Yu, Z.~Liang, and J.~Ma,
  ``Sparse-view {X}-ray {CT} reconstruction via total generalized variation
  regularization,'' \emph{Physics in medicine and biology}, vol.~59, no.~12, p.
  2997, 2014.

\bibitem{danielyan2011bm3d}
A.~Danielyan, V.~Katkovnik, and K.~Egiazarian, ``{BM3D} frames and variational
  image deblurring,'' \emph{IEEE Transactions on Image Processing}, vol.~21,
  no.~4, pp. 1715--1728, 2011.

\bibitem{elad2006dict}
M.~Elad and M.~Aharon, ``Image denoising via sparse and redundant
  representations over learned dictionaries,'' \emph{IEEE Transactions on Image
  Processing}, vol.~15, no.~12, pp. 3736--3745, 2006.

\bibitem{xu2020limited}
M.~Xu, D.~Hu, F.~Luo, F.~Liu, S.~Wang, and W.~Wu, ``Limited-angle x-ray ct
  reconstruction using image gradient $l_0$-norm with dictionary learning,''
  \emph{IEEE Transactions on Radiation and Plasma Medical Sciences}, vol.~5,
  no.~1, pp. 78--87, 2020.

\bibitem{wang2020deep}
G.~Wang, J.~C. Ye, and B.~De~Man, ``Deep learning for tomographic image
  reconstruction,'' \emph{Nature Machine Intelligence}, vol.~2, no.~12, pp.
  737--748, 2020.

\bibitem{wang2023review}
T.~Wang, W.~Xia, J.~Lu, and Y.~Zhang, ``A review of deep learning ct
  reconstruction from incomplete projection data,'' \emph{IEEE Transactions on
  Radiation and Plasma Medical Sciences}, 2023.

\bibitem{ongie2020deep}
G.~Ongie, A.~Jalal, C.~A. Metzler, R.~G. Baraniuk, A.~G. Dimakis, and
  R.~Willett, ``Deep learning techniques for inverse problems in imaging,''
  \emph{IEEE Journal on Selected Areas in Information Theory}, vol.~1, no.~1,
  pp. 39--56, 2020.

\bibitem{lucas2018}
A.~Lucas, M.~Iliadis, R.~Molina, and A.~K. Katsaggelos, ``Using deep neural
  networks for inverse problems in imaging: Beyond analytical methods,''
  \emph{IEEE Signal Processing Magazine}, vol.~35, no.~1, pp. 20--36, 2018.

\bibitem{zhang2017beyond}
K.~Zhang, W.~Zuo, Y.~Chen, D.~Meng, and L.~Zhang, ``Beyond a gaussian denoiser:
  Residual learning of deep cnn for image denoising,'' \emph{IEEE transactions
  on image processing}, vol.~26, no.~7, pp. 3142--3155, 2017.

\bibitem{chen2017low}
H.~Chen, Y.~Zhang, M.~K. Kalra, F.~Lin, Y.~Chen, P.~Liao, J.~Zhou, and G.~Wang,
  ``Low-dose {CT} with a residual encoder-decoder convolutional neural
  network,'' \emph{IEEE transactions on medical imaging}, vol.~36, no.~12, pp.
  2524--2535, 2017.

\bibitem{kang2017deep}
E.~Kang, J.~Min, and J.~C. Ye, ``A deep convolutional neural network using
  directional wavelets for low-dose x-ray ct reconstruction,'' \emph{Medical
  physics}, vol.~44, no.~10, pp. e360--e375, 2017.

\bibitem{huang2022deep}
Z.~Huang, Z.~Chen, G.~Quan, Y.~Du, Y.~Yang, X.~Liu, H.~Zheng, D.~Liang, and
  Z.~Hu, ``Deep cascade residual networks (dcrns): Optimizing an
  encoder--decoder convolutional neural network for low-dose ct imaging,''
  \emph{IEEE Transactions on Radiation and Plasma Medical Sciences}, vol.~6,
  no.~8, pp. 829--840, 2022.

\bibitem{zeng2021noise}
D.~Zeng, L.~Wang, M.~Geng, S.~Li, Y.~Deng, Q.~Xie, D.~Li, H.~Zhang, Y.~Li,
  Z.~Xu \emph{et~al.}, ``Noise-generating-mechanism-driven unsupervised
  learning for low-dose ct sinogram recovery,'' \emph{IEEE Transactions on
  Radiation and Plasma Medical Sciences}, vol.~6, no.~4, pp. 404--414, 2021.

\bibitem{jeon2022mm}
S.-Y. Jeon, W.~Kim, and J.-H. Choi, ``Mm-net: Multiframe and multimask-based
  unsupervised deep denoising for low-dose computed tomography,'' \emph{IEEE
  Transactions on Radiation and Plasma Medical Sciences}, vol.~7, no.~3, pp.
  296--306, 2022.

\bibitem{lehtinen2018noise2noise}
J.~Lehtinen, J.~Munkberg, J.~Hasselgren, S.~Laine, T.~Karras, M.~Aittala, and
  T.~Aila, ``Noise2noise: Learning image restoration without clean data,'' in
  \emph{International Conference on Machine Learning}.\hskip 1em plus 0.5em
  minus 0.4em\relax PMLR, 2018, pp. 2965--2974.

\bibitem{batson2019noise2self}
J.~Batson and L.~Royer, ``Noise2self: Blind denoising by self-supervision,'' in
  \emph{International Conference on Machine Learning}.\hskip 1em plus 0.5em
  minus 0.4em\relax PMLR, 2019, pp. 524--533.

\bibitem{krull2019noise2void}
A.~Krull, T.-O. Buchholz, and F.~Jug, ``Noise2void-learning denoising from
  single noisy images,'' in \emph{Proceedings of the IEEE/CVF Conference on
  Computer Vision and Pattern Recognition}, 2019, pp. 2129--2137.

\bibitem{laine2019high}
S.~Laine, T.~Karras, J.~Lehtinen, and T.~Aila, ``High-quality self-supervised
  deep image denoising,'' \emph{Advances in Neural Information Processing
  Systems}, vol.~32, 2019.

\bibitem{hendriksen2020noise2inverse}
A.~A. Hendriksen, D.~M. Pelt, and K.~J. Batenburg, ``Noise2inverse:
  Self-supervised deep convolutional denoising for tomography,'' \emph{IEEE
  Transactions on Computational Imaging}, vol.~6, pp. 1320--1335, 2020.

\bibitem{chen2021equivariant}
D.~Chen, J.~Tachella, and M.~E. Davies, ``Equivariant imaging: Learning beyond
  the range space,'' in \emph{Proceedings of the IEEE/CVF International
  Conference on Computer Vision}, 2021, pp. 4379--4388.

\bibitem{chen2022robust}
------, ``Robust equivariant imaging: a fully unsupervised framework for
  learning to image from noisy and partial measurements,'' in \emph{Proceedings
  of the IEEE/CVF Conference on Computer Vision and Pattern Recognition}, 2022,
  pp. 5647--5656.

\bibitem{dangovski2021equivariant}
R.~Dangovski, L.~Jing, C.~Loh, S.~Han, A.~Srivastava, B.~Cheung, P.~Agrawal,
  and M.~Soljacic, ``Equivariant self-supervised learning: Encouraging
  equivariance in representations,'' in \emph{International Conference on
  Learning Representations}, 2021.

\bibitem{buzug2009computed}
T.~M. Buzug, ``Computed tomography: from photon statistics to modern cone-beam
  ct,'' 2009.

\bibitem{hansen2021computed}
P.~C. Hansen, J.~J{\o}rgensen, and W.~R. Lionheart, \emph{Computed tomography:
  algorithms, insight, and just enough theory}.\hskip 1em plus 0.5em minus
  0.4em\relax SIAM, 2021.

\bibitem{schwab2019deep}
J.~Schwab, S.~Antholzer, and M.~Haltmeier, ``Deep null space learning for
  inverse problems: convergence analysis and rates,'' \emph{Inverse Problems},
  vol.~35, no.~2, p. 025008, 2019.

\bibitem{hastie2009elements}
T.~Hastie, R.~Tibshirani, J.~H. Friedman, and J.~H. Friedman, \emph{The
  elements of statistical learning: data mining, inference, and
  prediction}.\hskip 1em plus 0.5em minus 0.4em\relax Springer, 2009, vol.~2.

\bibitem{cohen2016group}
T.~Cohen and M.~Welling, ``Group equivariant convolutional networks,'' in
  \emph{International conference on machine learning}.\hskip 1em plus 0.5em
  minus 0.4em\relax PMLR, 2016, pp. 2990--2999.

\bibitem{lenc2015understanding}
K.~Lenc and A.~Vedaldi, ``Understanding image representations by measuring
  their equivariance and equivalence,'' in \emph{Proceedings of the IEEE
  conference on computer vision and pattern recognition}, 2015, pp. 991--999.

\bibitem{mohan2019robust}
S.~Mohan, Z.~Kadkhodaie, E.~P. Simoncelli, and C.~Fernandez-Granda, ``Robust
  and interpretable blind image denoising via bias-free convolutional neural
  networks,'' \emph{arXiv preprint arXiv:1906.05478}, 2019.

\bibitem{ioffe2015batch}
S.~Ioffe and C.~Szegedy, ``Batch normalization: Accelerating deep network
  training by reducing internal covariate shift,'' in \emph{International
  conference on machine learning}.\hskip 1em plus 0.5em minus 0.4em\relax PMLR,
  2015, pp. 448--456.

\bibitem{kingma2014adam}
D.~P. Kingma and J.~Ba, ``Adam: A method for stochastic optimization,''
  \emph{arXiv preprint arXiv:1412.6980}, 2014.

\bibitem{moen2021low}
T.~R. Moen, B.~Chen, D.~R. Holmes~III, X.~Duan, Z.~Yu, L.~Yu, S.~Leng, J.~G.
  Fletcher, and C.~H. McCollough, ``Low-dose ct image and projection dataset,''
  \emph{Medical physics}, vol.~48, no.~2, pp. 902--911, 2021.

\bibitem{van2015astra}
W.~Van~Aarle, W.~J. Palenstijn, J.~De~Beenhouwer, T.~Altantzis, S.~Bals, K.~J.
  Batenburg, and J.~Sijbers, ``The astra toolbox: A platform for advanced
  algorithm development in electron tomography,'' \emph{Ultramicroscopy}, vol.
  157, pp. 35--47, 2015.

\bibitem{jin2017deep}
K.~H. Jin, M.~T. McCann, E.~Froustey, and M.~Unser, ``Deep convolutional neural
  network for inverse problems in imaging,'' \emph{IEEE transactions on image
  processing}, vol.~26, no.~9, pp. 4509--4522, 2017.

\bibitem{song2021solving}
Y.~Song, L.~Shen, L.~Xing, and S.~Ermon, ``Solving inverse problems in medical
  imaging with score-based generative models,'' in \emph{International
  Conference on Learning Representations (ICLR)}, 2022.

\bibitem{wagner2022dual}
F.~Wagner, M.~Thies, L.~Pfaff, O.~Aust, S.~Pechmann, N.~Maul, M.~Rohleder,
  M.~Gu, J.~Utz, F.~Denzinger, and A.~Maier, ``On the benefit of dual-domain
  denoising in a self-supervised low-dose ct setting,'' \emph{arXiv preprint
  arXiv:2211.01111}, 2022.

\bibitem{liu2021self}
X.~Liu, F.~Zhang, Z.~Hou, L.~Mian, Z.~Wang, J.~Zhang, and J.~Tang,
  ``Self-supervised learning: Generative or contrastive,'' \emph{IEEE
  Transactions on Knowledge and Data Engineering}, 2021.

\bibitem{desai2021vortex}
A.~D. Desai, B.~Gunel, B.~M. Ozturkler, H.~Beg, S.~Vasanawala, B.~A.
  Hargreaves, C.~R{\'e}, J.~M. Pauly, and A.~S. Chaudhari, ``Vortex:
  Physics-driven data augmentations using consistency training for robust
  accelerated mri reconstruction,'' in \emph{International Conference on
  Medical Imaging with Deep Learning}.\hskip 1em plus 0.5em minus 0.4em\relax
  PMLR, 2022, pp. 325--352.

\bibitem{choi2023self}
K.~Choi, S.~H. Kim, and S.~Kim, ``Self-supervised denoising of projection data
  for low-dose cone-beam ct,'' \emph{Medical Physics}, 2023.

\end{thebibliography}

\end{document}